\documentclass[12pt]{iopart}
\usepackage{iopams,mathrsfs,graphicx,hyperref,color,bm,lipsum}
\expandafter\let\csname equation*\endcsname\relax
\expandafter\let\csname endequation*\endcsname\relax
\newcommand{\gae}{\lower 2pt \hbox{$\, \buildrel {\scriptstyle >}\over {\scriptstyle \sim}\,$}}
\newcommand{\lae}{\lower 2pt \hbox{$\, \buildrel {\scriptstyle <}\over {\scriptstyle \sim}\,$}}
\usepackage{amsmath}

\DeclareMathOperator{\arccosh}{acosh}
\newcommand{\ee}{\mathrm{e}}
\newcommand{\ii}{\mathrm{i}}

\newcommand{\dd}[1]{{\mathrm{d}} #1~}

\DeclareRobustCommand{\vect}[1]{\bm{#1}}
\begin{document}

\title[Discrete space-time resetting model]{Discrete space-time resetting model: Application to first-passage and transmission statistics}
\author{Debraj Das}
\address{Department of Engineering Mathematics, University of
Bristol, Bristol BS8 1TW, United Kingdom}
\ead{debrajdasphys@gmail.com, debraj.das@bristol.ac.uk}

\author{Luca Giuggioli}
\address{Bristol Centre for Complexity Sciences and Department of Engineering Mathematics, University of
Bristol, Bristol BS8 1TW, United Kingdom}
\ead{Luca.Giuggioli@bristol.ac.uk}

\vspace{10pt}
\begin{indented}
\item[]\today
\end{indented}

\begin{abstract}
We consider the dynamics of lattice random walks with resetting. The walker moving randomly on a lattice of arbitrary dimensions resets at every time step to a given site with a constant probability $r$. We construct a discrete renewal equation and present closed-form expressions for different quantities of the resetting dynamics in terms of the underlying reset-free propagator or Green's function. We apply our formalism to the biased random walk dynamics in one-dimensional unbounded space and show how one recovers in the continuous limits results for diffusion with resetting. The resetting dynamics of biased random walker in one-dimensional domain bounded with periodic and reflecting boundaries is also analyzed. Depending on the bias the first-passage probability in periodic domain shows multi-fold non-monotonicity as $r$ is varied. Finally, we apply our formalism to study the transmission dynamics of two lattice walkers with resetting in one-dimensional domain bounded by periodic and reflecting boundaries. The probability of a definite transmission between the walkers shows non-monotonic behavior as the resetting probabilities are varied.
\end{abstract}


\section{Introduction}
Dynamical processes subject to stochastic resetting are observed in a wide range of natural phenomena. For example, in population dynamics a sudden catastrophe may reduce the size of a microbial population drastically to a previous value~\cite{visco_switching_2010,manrubia_stochastic_1999}, in financial markets a random event may cause crash in stock prices and bring them to a much lower value~\cite{sornette_critical_2003}, in transcription process sudden interruptions due to RNA cleavage may cause backtrack recovery of polymerases~\cite{roldan_stochastic_2016}. Stochastic resetting is also important in search processes where the searching agents often spontaneously reset back to a fixed point after an unsuccessful trial, e.g., humans searching a string on the internet, animals foraging~\cite{viswanathan_physics_2011}, and biomolecules looking up for binding sites~\cite{coppey_kinetics_2004}. Stochastic resetting, which allows for long-range movements on top of an underlying local dynamics, often appears beneficial in optimizing search strategies~\cite{benichou_optimal_2005,benichou_intermittent_2007}, e.g., in computer science in finding an optimal completion time for an algorithm~\cite{luby_optimal_1993,tong_random_2008}.

`Stochastic resetting' or `resetting' was first considered in statistical physics by Evans and Majumdar in their seminal work~\cite{evans_diffusion_2011}, where the authors studied the diffusive dynamics of a Brownian particle subject to resetting with a constant rate to the initial position.
In this work, it has been shown that the resetting dynamics attains a nonequilibrium stationary state (NESS), and that the mean first-passage time (MFPT) has a minimum at an optimal value of the resetting rate.
Since then, many generalization of this resetting dynamics of Brownian particle have been studied: with spatially-dependent resetting rates~\cite{evans_diffusion_2011-1,roldan_path-integral_2017}, in arbitrary spatial dimensions~\cite{evans_diffusion_2014}, in presence of partially absorbing spatial point~\cite{whitehouse_effect_2013} or regions~\cite{schumm_search_2021}, in bounded domains~\cite{christou_diffusion_2015, chatterjee_diffusion_2018}, in presence of potential~\cite{pal_diffusion_2015,singh_resetting_2020}, with different choices of resetting position~\cite{evans_diffusion_2011-1,boyer_random_2014,majumdar_random_2015}, etc.
Resetting with a constant rate implies that the random waiting times between two successive resetting events are sampled from an exponential distribution, and therefore often referred to Poissonian resetting~\cite{evans_stochastic_2020}. Non-Poissonian resetting with different waiting time distributions have also been studied~\cite{eule_non-equilibrium_2016,pal_diffusion_2016,nagar_diffusion_2016}.
The canonical diffusion dynamics with resetting triggered when the diffusing particle reaches a predetermined threshold shows a non-stationary behavior~\cite{de_bruyne_optimization_2020}. 
In search problems that are usually studied as first-passage processes, the nonequilibrium resetting dynamics offers more efficient strategy than an equivalent equilibrium dynamics~\cite{evans_optimal_2013,giuggioli_comparison_2019,pal_search_2020}.
The optimal mean time needed by a freely diffusing particle to reach a target in presence of resetting has been studied experimentally and theoretically~\cite{besga_optimal_2020,faisant_optimal_2021}. In another experimental realization of resetting, it is shown that the energetic cost of resetting in first-passage scenarios cannot be made arbitrarily small~\cite{tal-friedman_experimental_2020}.
Resetting has also been studied in interacting particle systems such as fluctuating surfaces~\cite{gupta_fluctuating_2014,majumdar_dynamical_2015,gupta_resetting_2016}
 and coagulation-diffusion process~\cite{durang_statistical_2014}, in non-diffusive processes such as L\'{e}vy flights~\cite{kusmierz_first_2014,kusmierz_optimal_2015} and run and tumble dynamics~\cite{evans_run_2018}, in telegraphic process~\cite{masoliver_telegraphic_2019}, in active Brownian process~\cite{kumar_active_2020},
in symmetric exclusion process~\cite{basu_symmetric_2019},
in Ising model~\cite{magoni_ising_2020}, as well as in quantum systems~\cite{mukherjee_quantum_2018,das_quantum_2022}. 
In stochastic thermodynamics, the first and the second law of resetting are derived and the contribution to entropy production due to resetting has also been identified~\cite{fuchs_stochastic_2016}.   

In the area of random walks with resetting, the literature is arguably dominated with studies that have considered continuous time. However, there are recent works that address stochastic processes with resetting in discrete times~\cite{kusmierz_first_2014, boyer_random_2014, majumdar_random_2015, falcon-cortes_localization_2017, boyer_anderson-like_2019, riascos_random_2020, bonomo_first_2021, biroli_number_2022}. 
In Ref.~\cite{kusmierz_first_2014}, the discrete-time resetting dynamics of a random walker in space with step-lengths sampled from a continuous distribution was considered. A non-Markovian discrete-time random walk on a lattice with preferential resetting to previously visited sites has been addressed in Ref.~\cite{boyer_random_2014}, and is further studied with partially trapping inhomogeneities in Refs.~\cite{falcon-cortes_localization_2017, boyer_anderson-like_2019}. Another non-Markovian resetting dynamics in lattice random walk, where the walker remembers the maximum location visited so far and resets to that site, has been studied in Ref.~\cite{majumdar_random_2015}. Reference~\cite{riascos_random_2020} has addressed the lattice random walk dynamics with resetting to the initial position in arbitrary complex networks, while Ref.~\cite{biroli_number_2022} investigates the number of distinct sites visited by a resetting random walker on hypercubic lattice. 

First-passage process on lattices with resetting in discrete times has been studied in Ref.~\cite{bonomo_first_2021}. By generalizing a completion process approach from Ref.~\cite{pal_first_2017} to discrete resetting, the first-passage probability and its moments have been derived in terms of quantities without resetting. The formalism has been applied to the symmetric and biased random walker dynamics with geometrical and sharp~\cite{pal_diffusion_2016,pal_first_2017,eliazar_mean-performance_2020} resetting in one-dimensional (1D) domain bounded by two absorbing boundaries.
In this work, we tackle the biased random walk problem with resetting by using a different approach.
We construct a discrete spatio-temporal renewal equation for the lattice-walk probability with resetting and express different quantities in terms of the reset-free propagator.
Specifically, we obtain closed-form expressions for the resetting propagator, the stationary-state probability, the first-passage probability, the return probability, the MFPT, and the mean return time. 
Additionally, in presence of a partially absorbing site, we present explicit analytical formulae for the survival probability, the absorption probability, and the mean absorption time.
Our formalism is quite general and may be applied to any underlying Markovian stochastic process (in the absence of resetting). 
As a demonstration, we apply our formalism to the dynamics of a biased lazy random walker (BLRW) in 1D unbounded domain and domains bounded by periodic and reflecting boundaries. 
We also show how in the continuous spatio-temporal limits one recovers corresponding results for diffusion with resetting.
We then apply our formalism to the transmission statistics of two 1D BLRWs with resetting in bounded domain. This is done by studying the multi-target first-passage statistics of a single 2D BLRW in bounded domain with partially absorbing sites.

The paper is organized as follows. In Sec.~\ref{sec:TheResettingModel}, we introduce the discrete space-time resetting model and show how one may express various quantities of interest in terms of the reset-free propagator. Section~{\ref{subsec:prop-unbounded}}\, is devoted to the application of our formalism to the BLRW dynamics with resetting in 1D unbounded domain, while Sec.~\ref{subsec:prop-bounded}\, deals with that in domains bounded by periodic and reflecting boundaries.
In Sec.~\ref{sec:trans}, we study the transmission statistics between two 1D BLRWs with resetting in bounded domain. Here, we consider two different resetting schemes of the individual 1D walkers: (i) simultaneous resetting in Sec.~\ref{subsec:simul-reset}, and (ii) independent resetting in Sec.~\ref{subsec:indep-reset}. The study of transmission events between the two 1D BLRWs is taken up in Sec.~\ref{subsec:trans-process}. 
We draw our conclusions in Sec.~\ref{sec:conclu}.

\section{The discrete resetting model}
\label{sec:TheResettingModel}
We consider the discrete-time dynamics of a random walker subject to stochastic resetting in a $d$-dimensional discrete lattice space. The position of every lattice site is given by a $d$-component vector  $\bm{n}$, such that $\bm{n}=(n_1,n_2,\ldots,n_d)$ with $n_i$ an integer for all $i \in [1,2,\ldots,d\,]$.
The dynamics of the walker is the following. Starting from a site $\bm{n}_0$ at time $t = 0$, the walker at every subsequent time step either resets to a given site $\bm{n}_c$ with a given probability $r$ with $0\leq r\leq 1$, or performs its underlying Markov walk with the
complementary probability $(1-r)$. 
For this resetting dynamics, the propagator for the walker, i.e., the probability to find the walker at a site $\bm{n}$
at time $t$ while starting from site $\bm{n}_0$ at time $t=0$ is denoted by $Q(\bm{n},t|\bm{n}_0)$. The quantity $Q(\bm{n},t|\bm{n}_0)$ may be expressed in a renewal form in terms of the reset-free propagator, denoted by $Q_{\mathrm{NR}}(\bm{n},t|\bm{n}_0)$ as follows (see~\ref{app:derive_renewal_eq}\, for the derivation)
\begin{align}
Q(\bm{n},t|\bm{n}_0) =r\sum_{t'=0}^{t-1} (1-r)^{t'} Q_{\mathrm{NR}}(\bm{n},t'|\bm{n}_c)+(1-r)^t Q_{\mathrm{NR}}(\bm{n},t|\bm{n}_0) \, . \label{eq:reset-x}
\end{align} 
The aspect that Eq.~\eqref{eq:reset-x} directly connects the resetting propagator $Q(\bm{n},t|\bm{n}_0)$ to the reset-free propagator $Q_{\mathrm{NR}}(\bm{n},t|\bm{n}_0)$ is particularly convenient as it allows us to express various quantities of the resetting dynamics in terms of the reset-free propagator.

From Eq.~\eqref{eq:reset-x}, one may obtain the renewal equation in continuous space ($\bm{x}$) and continuous time $(\tau)$ by using
$\bm{x}=\bm{n}L$ and $\tau=t T$, and considering appropriate limits. Here,  $L$ is the length of the lattice spacing, and $T$ is the time duration of a single step. To proceed, one has to take the simultaneous limits $\bm{n} \to \infty, \, L \to 0$ keeping $\bm{x}$ fixed, and the simultaneous
limits $t \to \infty, \, T \to 0$ keeping $\tau$ fixed. 
The continuous-time resetting rate $R$ is related to the discrete-time resetting probability $r$ by the relation $R=r/T$. Then, to keep $R$ fixed and finite, one needs to consider the simultaneous limits $T\to 0, ~r\to 0$. In these limits, we get
$(1-r)^t=\exp[t \log(1-r)]=\exp[(\tau/T)\log(1-RT)] \to \exp(-R\tau)$.
On the other hand, noting that $\Delta t'=1$ for discrete time steps, one may write
$\sum_{t'=0}^{t-1} r \, (1-r)^{t'} = \sum_{t'=0}^{t-1} \Delta t' \, r \,(1-r)^{t'} = \sum_{\tau'=0}^{\tau-T} \,\Delta \tau' \, R \,\exp[(\tau'/T)\log(1-RT)]  $. In the limit $T\to 0$, i.e., the limit $\Delta \tau' = \Delta t'\, T = T \to 0$, we have $\sum_{t'=0}^{t-1} r(1-r)^{t'} \to \int_{0}^{\tau}\dd{\tau'} R \, \exp(-R\tau')$.
In these limits, the renewal equation~\eqref{eq:reset-x} transforms into the well-known continuous space-time version~\cite{gupta_fluctuating_2014,giuggioli_comparison_2019,evans_stochastic_2020}
\begin{align}
P(\bm{x},\tau| \bm{x}_0)= R \int_0^\tau \dd{\tau'} \ee^{-R \tau'} \, P_{\mathrm{NR}}(\bm{x}, \tau' |\bm{x}_c)+\ee^{-R \tau} \, P_{\mathrm{NR}}(\bm{x},\tau|\bm{x}_0) \, ,
\label{eq:reset-continuum}
\end{align}
where $\bm{x}_c$ is the resetting point in the $d$-dimensional continuous space, and the quantities $P(\bm{x}, \tau |\bm{x}')$ and $P_{\mathrm{NR}}(\bm{x}, \tau |\bm{x}')$ are respectively the propagators with and without resetting in continuous space-time.

\subsection{Propagator in $z$-domain}
We define the generating function $\widetilde{Q}(\bm{n},z|\bm{n}_0)$ for the time-dependent propagator $Q(\bm{n},t|\bm{n}_0)$ with resetting, i.e., $\widetilde{f}(z) \equiv \sum_{t=0}^{\infty} \, z^t \, f(t)$ for any time-dependent function $f(t)$.
In $z$-domain, one may then rewrite the renewal equation~\eqref{eq:reset-x} as
\begin{align}
\widetilde{Q}(\bm{n},z|\bm{n}_0) = \frac{r z}{1-z} ~ \widetilde{Q}_{\mathrm{NR}}(\bm{n},(1-r)z|\bm{n}_c) +  \widetilde{Q}_{\mathrm{NR}}(\bm{n},(1-r)z|\bm{n}_0) \, , \label{eq:Qreset_nz}
\end{align}
where $\widetilde{Q}_{\mathrm{NR}}(\bm{n},z|\bm{n}_0) $ is the generating function for the time-dependent reset-free propagator $Q_{\mathrm{NR}}(\bm{n},t|\bm{n}_0)$.

\subsection{Stationary-state probability}
In the limit $t \to \infty$, using the final value theorem for the $z$-transform, from Eq.~\eqref{eq:Qreset_nz} we obtain the stationary-state probability  with resetting  ${}^{\mathrm{ss}}{Q}(\bm{n}) \equiv \lim_{t \to \infty } {Q}(\bm{n}, t  |\bm{n}_0) = \lim_{z \to 1}[(1-z)\, \widetilde{Q}(\bm{n},z|\bm{n}_0) ] $ as
\begin{align}
{}^{\mathrm{ss}}{Q}(\bm{n}) =   r \,\widetilde{Q}_{\mathrm{NR}}(\bm{n},(1-r)|\bm{n}_c) \, . \label{eq:Qreset_SS}
\end{align}
Equation~\eqref{eq:Qreset_SS} clearly shows that the stationary-state probability for the resetting dynamics depends on the resetting site $\bm{n}_c$ and is independent of the initial site $\bm{n}_0$.

\subsection{First-passage processes}
The very first occurrence of an event is often an important feature of a stochastic dynamics. In our case, these events might be represented by (i) the random walker reaching a site $\bm{n}$ for the first time while having been started from a site $\bm{n}_0 \neq \vect{n}$, or (ii) the random walker returning to a site $\bm{n}$ for the first time while having been started from the same site $\bm{n}$. We refer the associated probability of occurrence of these two events to the first-passage probability denoted by ${\mathcal F}(\bm{n}, t | \bm{n}_0)$ and the first-return probability denoted by ${\mathcal R}(\bm{n}, t)$, respectively. 
Both of these probabilities are linked with the propagator via a renewal relation~\cite{montroll_random_1965,hughes_random_1995,redner_guide_2001}, which in  $z$-domain yields 
$\widetilde{\mathcal F}(\bm{n},z|\bm{n}_0) = \widetilde{Q}(\bm{n},z|\bm{n}_0)/\widetilde{Q}(\bm{n},z|\bm{n}) $  
and $ \widetilde{\mathcal R}(\bm{n},z) = 1 - 1/  \widetilde{Q}(\bm{n},z|\bm{n}) $. As these relations hold for Markov processes with and without resetting, using Eq.~\eqref{eq:Qreset_nz}, we obtain the generating functions $\widetilde{\mathcal  F}(\bm{n},z|\bm{n}_0)$ and $\widetilde{\mathcal  R}(\bm{n},z)$ for the resetting dynamics in terms of the generating function of the reset-free propagator as
\begin{align}
\widetilde{\mathcal F}(\bm{n},z|\bm{n}_0)  
= \frac{ r z \, \widetilde{Q}_{\mathrm{NR}}(\bm{n},(1-r)z|\bm{n}_c) + (1-z)  \widetilde{Q}_{\mathrm{NR}}(\bm{n},(1-r)z|\bm{n}_0) }{r z \, \widetilde{Q}_{\mathrm{NR}}(\bm{n},(1-r)z|\bm{n}_c) + (1-z)  \widetilde{Q}_{\mathrm{NR}}(\bm{n},(1-r)z|\bm{n})  }   \, , \label{eq:Qreset_FP'}
\end{align}
and
\begin{align}
\widetilde{\mathcal  R}(\bm{n},z) &=  \frac{ r z~ \widetilde{Q}_{\mathrm{NR}}(\bm{n},(1-r)z|\bm{n}_c) + (1-z) \Big[ \widetilde{Q}_{\mathrm{NR}}(\bm{n},(1-r)z|\bm{n}) - 1  \Big]}{ r z~ \widetilde{Q}_{\mathrm{NR}}(\bm{n},(1-r)z|\bm{n}_c) + (1-z)  \widetilde{Q}_{\mathrm{NR}}(\bm{n},(1-r)z|\bm{n})  } \, , \label{eq:Qreset_ReturnP}
\end{align}
respectively.

The right-hand side of Eq.~\eqref{eq:Qreset_FP'} may also be expressed in terms of the generating function of the reset-free first-passage probability $\widetilde{\mathcal{F}}_{\mathrm{NR}}(\bm{n},z|\bm{n}_0) \equiv \widetilde{Q}_{\mathrm{NR}}({\vect n},z|{\vect n}_0) / \widetilde{Q}_{\mathrm{NR}}({\vect n},z|{\vect n})$. 
Then, we may write $\widetilde{\mathcal F}(\bm{n},z|\bm{n}_0) =  \big[ r z  \widetilde{\mathcal{F}}_{\mathrm{NR}}(\bm{n},(1-r)z|\bm{n}_c) + (1-z)  \widetilde{\mathcal{F}}_{\mathrm{NR}}(\bm{n},(1-r)z|\bm{n}_0) \big]  / \big[ r z   \widetilde{\mathcal{F}}_{\mathrm{NR}}(\bm{n},(1-r)z|\bm{n}_c) + (1-z)  \big] $. 
To make contact with Ref.~\cite{bonomo_first_2021}, we note that if one considers that resetting can also occur at time $t=0$ with probability $r$, then the renewal equation~\eqref{eq:reset-x} takes the form 
$Q(\bm{n},t|\bm{n}_0) =r\sum_{t'=0}^{t} (1-r)^{t'} Q_{\mathrm{NR}}(\bm{n},t'|\bm{n}_c)+(1-r)^{t+1} Q_{\mathrm{NR}}(\bm{n},t|\bm{n}_0)$.
In this case, the generating function of the first-passage probability is given by $\widetilde{\mathcal F}(\bm{n},z|\bm{n}_0) =  \big[ r   \widetilde{\mathcal{F}}_{\mathrm{NR}}(\bm{n},(1-r)z|\bm{n}_c) + (1-r)(1-z)  \widetilde{\mathcal{F}}_{\mathrm{NR}}(\bm{n},(1-r)z|\bm{n}_0) \big] / \big[ r   \widetilde{\mathcal{F}}_{\mathrm{NR}}(\bm{n},(1-r)z|\bm{n}_c) + (1-r)(1-z)  \big]  $, 
which simplifies to Eq.~(23) in Ref.~\cite{bonomo_first_2021} when ${\vect{n}}_c={\vect{n}}_0$.

From Eqs.~\eqref{eq:Qreset_FP'} and~\eqref{eq:Qreset_ReturnP}, one may see that we have $\widetilde{\mathcal  F}(\bm{n},1|\bm{n}_0) = 1$ and $\widetilde{\mathcal  R}(\bm{n},1) = 1$, i.e., the walker reaches or returns to site $\vect{n}$ with probability one.
The certainty to reach or return to a site does not necessarily hold for a reset-free random-walk as popularized by P\'{o}lya in his seminal work~\cite{polya_quelques_1919,polya_uber_1921} on the return probability on Euclidean lattices of dimensions larger that two, but also in some instances for a reset-free biased walker in 1D unbounded domain (see~\ref{sec:FP-RF-BLRW}). 
Thus, in general, although the reset-free quantities $\widetilde{\mathcal  F}_{\mathrm{NR}}(\bm{n},1|\bm{n}_0) $ and $\widetilde{\mathcal  R}_{\mathrm{NR}}(\bm{n},1) $ do depend on the functional form of $\widetilde{Q}_{\mathrm{NR}}(\bm{n},z|\bm{n}_0)$, for the resetting case $\widetilde{\mathcal  F}(\bm{n},1|\bm{n}_0)$ and $\widetilde{\mathcal R}(\bm{n},1)$ do not.

The MFPT $T_{\bm{n}_0 \to \bm{n}}$ to reach site $\bm{n}$ starting from site $\bm{n}_0$ obtained from the generating function $\widetilde{\mathcal  F}(\bm{n},z|\bm{n}_0)$ using the usual relation $ T_{\bm{n}_0 \to \bm{n}} \equiv \big[\frac{\partial}{\partial z} \widetilde{ \mathcal  F}(\bm{n},z|\bm{n}_0) \big]_{z=1}$ and similarly the mean first-return time 
($ R_{\bm{n}} \equiv \big[\frac{\partial}{\partial z} \widetilde{\mathcal  R}(\bm{n},z) \big]_{z=1}$) are given, respectively, by
\begin{align}
T_{\bm{n}_0 \to \bm{n}} = \frac{\widetilde{Q}_{\mathrm{NR}}(\bm{n},(1-r)|\bm{n})-\widetilde{Q}_{\mathrm{NR}}(\bm{n},(1-r)|\bm{n}_0)}{r \,\widetilde{Q}_{\mathrm{NR}}(\bm{n},(1-r)|\bm{n}_c)}   \, \label{eq:Qreset_MFPT'} 
\end{align}
and
\begin{align}
R_{\bm{n}} = \frac{1}{r \,\widetilde{Q}_{\mathrm{NR}}(\bm{n},(1-r)|\bm{n}_c)}  = \frac{1}{{}^{\mathrm{ss}}{Q}(\bm{n})}\, , \label{eq:Qreset_MRT'} 
\end{align}
as expected from Kac's Lemma~\cite{kac_notion_1947}.

\subsection{Dynamics with defect}
For many stochastic processes, there exists special spatial locations, where the underlying dynamics are different from the rest of the space. These special locations, customarily referred to defects, may be partially absorbing traps~\cite{kenkre_theory_1980,den_hollander_random_1982}, or sites with  different hopping probabilities to the neighboring sites~\cite{kenkre_exciton_1982,kenkre_molecular_2008}. For our resetting dynamics, defect is defined as a partially absorbing trap at site $\vect{n}_{\mathrm d}$ with absorption probability $\rho ~( 0 \leq \rho \leq 1)$, with $\rho =1 $ corresponding to perfect absorption. We denote the generating function of the resetting propagator with the defect by $\widetilde{\mathcal{P}}(\bm{n}, z| \bm{n}_0)$. Using the so-called Montroll's defect technique~\cite{montroll_random_1965,montroll_effect_1955,montroll_chapter_1979,kenkre_memory_2021}, one obtains an exact solution for $\widetilde{\mathcal{P}}(\bm{n}, z| \bm{n}_0)$ in terms of the defect-free quantity $\widetilde{Q}(\bm{n}, z| \,\bm{n}_0)$, namely,
$\widetilde{\mathcal{P}}(\bm{n}, z| \bm{n}_0) =  \widetilde{Q}(\bm{n}, z| \,\bm{n}_0)    - \widetilde{Q}(\bm{n}, z| {\vect n}_{\mathrm d}) \, \widetilde{Q}(\bm{n}_{\mathrm d}, z| \,\bm{n}_0) / [ 1/\rho  - 1 + \widetilde{Q}(\bm{n}_{\mathrm d}, z| {\vect n}_{\mathrm d}) ] $. 
In presence of the partially absorbing trap, the quantity $\widetilde{\mathcal{P}}(\bm{n}, z| \bm{n}_0)$ is no longer normalized as may be seen from the sum $\sum_{\vect{n}} \widetilde{\mathcal{P}}(\bm{n}, z| \bm{n}_0)$. 
This sum rather yields the generating function of the survival probability $S(t|\vect{n}_0)$, i.e., the probability of not having been absorbed until time $t$ while starting from $\vect{n}_0$. In $z$-domain, we then have $\widetilde{S}(z| \vect{n}_0) = \sum_{\vect{n}} \widetilde{\mathcal{P}}(\bm{n}, z| \bm{n}_0) =
 [ 1 - \widetilde{Q}(\bm{n}_{\mathrm d}, z| {\vect n}_0) / \{ 1/\rho - 1 + \widetilde{Q}(\bm{n}_{\mathrm d}, z| {\vect n}_{\mathrm d}) \} ]/(1-z) \, ,
$
which with Eq.~\eqref{eq:Qreset_nz} gives
\begin{align}
\widetilde{S}(z| \vect{n}_0) &=  \frac{ \frac{1}{\rho} - 1 +  \widetilde{Q}_{\mathrm{NR}}(\bm{n}_{\mathrm d}, (1-r)z| {\vect n}_{\mathrm d}) - \widetilde{Q}_{\mathrm{NR}}(\bm{n}_{\mathrm d}, (1-r)z| {\vect n}_{0}) } {\left(\frac{1}{\rho} - 1\right) \! (1-z) + r   z \, \widetilde{Q}_{\mathrm{NR}}(\bm{n}_{\mathrm d},(1-r)z|\bm{n}_c) + (1-z)  \widetilde{Q}_{\mathrm{NR}}(\bm{n}_{\mathrm d},(1-r)z|\bm{n}_{\mathrm d})}  \, . \label{eq:Qreset_SP_in_reset_free_SP}
\end{align}

To connect to the available literature, one may also rewrite Eq.~\eqref{eq:Qreset_SP_in_reset_free_SP} in terms of the generating function of the reset-free survival probability $\widetilde{S}_{\mathrm{NR}}(z| \vect{n}_0) \equiv [ 1 - \widetilde{Q}_{\mathrm{NR}}(\bm{n}_{\mathrm d}, z| {\vect n}_0) / \{ 1/\rho - 1 + \widetilde{Q}_{\mathrm{NR}}(\bm{n}_{\mathrm d}, z| {\vect n}_{\mathrm d}) \} ]/(1-z)$  as
\begin{align}
\widetilde{S}(z| \vect{n}_0) =  \frac{ \widetilde{S}_{\mathrm{NR}} \big( (1-r)z | {\vect n}_0 \big) }{  1-r z \, \widetilde{S}_{\mathrm{NR}} \big( (1-r)z | {\vect n}_c \big)   }\, . \label{eq:Qreset_SP_in_reset_free_SP_NR}
\end{align}
Equation~\eqref{eq:Qreset_SP_in_reset_free_SP_NR} when ${\vect{n}}_c={\vect{n}}_0$ reduces to the general result Eq.~(5) in Ref.~\cite{kusmierz_first_2014}, where it represents the 
survival probability for a resetting walker moving on a real line that does not cross the origin. Our Eqs.~\eqref{eq:Qreset_SP_in_reset_free_SP} and~\eqref{eq:Qreset_SP_in_reset_free_SP_NR} generalize that result for partially absorbing targets.

The absorption probability ${{\mathcal{A}} (\vect{n}_{\mathrm d},t| \vect{n}_0)}$, i.e., the probability of the walker to get absorbed at time $t$ at site $\vect{n}_{\mathrm d}$ starting from site $\vect{n}_{0}$ is related to the survival probability $S(t| \vect{n}_0)$ by ${{\mathcal{A}} (\vect{n}_{\mathrm d},t| \vect{n}_0)} = S(t-1| \vect{n}_0) - S(t| \vect{n}_0)$. 
The generating function of the absorption probability, ${\widetilde{\mathcal{A}} (\vect{n}_{\mathrm d},z| \vect{n}_0)} = 1- (1-z) \widetilde{S}(z| \vect{n}_0)$,\cite{giuggioli_exact_2020} is given by
\begin{align}
{\widetilde{\mathcal{A}} (\vect{n}_{\mathrm d},z| \vect{n}_0)} & \nonumber \\ 
&\hskip-50pt = \frac{ r z \, \widetilde{Q}_{\mathrm{NR}}(\bm{n}_{\mathrm d},(1-r)z|\bm{n}_c) + (1-z)  \widetilde{Q}_{\mathrm{NR}}(\bm{n}_{\mathrm d},(1-r)z|\bm{n}_0) }{\left(\frac{1}{\rho} - 1\right) \! (1-z) + r   z \, \widetilde{Q}_{\mathrm{NR}}(\bm{n}_{\mathrm d},(1-r)z|\bm{n}_c) + (1-z)  \widetilde{Q}_{\mathrm{NR}}(\bm{n}_{\mathrm d},(1-r)z|\bm{n}_{\mathrm d})  }   \, , \label{eq:Qreset_AP'}
\end{align}
and the corresponding mean absorption time $ A_{\vect{n}_0 \to \vect{n}_{\mathrm d}} \equiv \big[\frac{\partial}{\partial z} {\widetilde{\mathcal{A}} (\vect{n}_{\mathrm d},z| \vect{n}_0)} \big]_{z=1}$ by
\begin{align}
A_{\bm{n}_0 \to \bm{n}_{\mathrm d}} =  \frac{ \widetilde{Q}_{\mathrm{NR}}(\bm{n}_{\mathrm d},(1-r)|\bm{n}_{\mathrm d})-\widetilde{Q}_{\mathrm{NR}}(\bm{n}_{\mathrm d},(1-r)|\bm{n}_0) + \frac{1}{\rho} - 1  }{r \,\widetilde{Q}_{\mathrm{NR}}(\bm{n}_{\mathrm d},(1-r)|\bm{n}_c)} \, . \label{eq:Qreset_MFAT'} 
\end{align}

From Eqs.~\eqref{eq:Qreset_MFPT'},~\eqref{eq:Qreset_MRT'}, and~\eqref{eq:Qreset_MFAT'}, one deduces the known result $A_{\bm{n}_0 \to \bm{n}_{\mathrm d}} = T_{\vect{n}_0 \to \vect{n}_{\mathrm d}} + (1/\rho - 1) R_{\vect{n}_\mathrm{d}}$~\cite{giuggioli_exact_2020}, which shows the existence of two temporal scales in presence of partially absorbing trap. The first term $T_{\vect{n}_0 \to \vect{n}_{\mathrm d}}$ is the MFPT to reach the defective site $\vect{n}_{\mathrm d}$, while the second $\rho$-dependent term gives a measure of the reaction time to get absorbed at that site, except when $\rho =1 $, which represents the motion limited~\cite{kenkre_exciton_1982} case as the walker gets absorbed as soon as it reaches the defective site. 
In the limit $\rho \to 0$, one has the reaction limited~\cite{kenkre_exciton_1982} case given by
$A_{\bm{n}_0 \to \bm{n}_{\mathrm d}} \simeq   R_{\vect{n}_\mathrm{d}} / \rho$.

We emphasize that Eqs.~\eqref{eq:Qreset_nz}--\eqref{eq:Qreset_SP_in_reset_free_SP},~\eqref{eq:Qreset_AP'}, and~\eqref{eq:Qreset_MFAT'} directly express various quantities for the resetting dynamics in terms of the generating function of reset-free propagator. Moreover, as no evolution rule is mentioned about the underlying reset-free dynamics, the formalism introduced in this section may be applied to any stochastic dynamics as long as it is Markovian.

\section{Resetting in one dimension}
\label{sec:prop1d}

We now apply the general formalism introduced in Section~\ref{sec:TheResettingModel} to the discrete-time dynamics of a BLRW moving on a 1D lattice in presence of stochastic resetting.
Recently, the dynamics of symmetric~\cite{giuggioli_exact_2020} and biased~\cite{sarvaharman_closed-form_2020} lattice walkers without resetting have been studied extensively. 
In the absence of resetting, the 1D BLRW dynamics is conveniently described by the following parameters~\cite{sarvaharman_closed-form_2020}: (i) the diffusion parameter $q$ with $0\leq q\leq 1$ and (ii) the bias parameter $g $ with $-1\leq g \leq 1$. Starting from a site $n$ at time $t$, the BLRW at the next instant $(t+1)$ either moves to site $(n-1)$ with probability $q(1+g)/2$, or to site $(n+1)$ with probability $q(1-g)/2$, or it stays at site $n$ with probability $(1-q)$. 
The dynamics is biased towards negative $x$-axis for $g>0$ and towards positive $x$-axis for $g<0$, while the magnitude $|g|$ controls the strength of the bias. 
For completeness, we have reported in~\ref{sec:BLRW}\, the necessary expressions taken from Ref.~\cite{sarvaharman_closed-form_2020}.

\subsection{Unbounded domain}
\label{subsec:prop-unbounded}
 
We first consider the resetting dynamics of the BLRW in 1D unbounded domain. In this case, the walker at every time step either resets to a given site $n_c$ with probability $r$, or follows its underlying dynamics described in the previous paragraph with probability $(1-r)$.

\subsubsection{Occupation probability}
\label{subsubsec:occu-prop-unbounded}
Substituting the expression for the quantity $\widetilde{Q}_{\mathrm{NR}}(n,z|n_0)$ from Eq.~\eqref{eq:defect-free-prop} in Eq.~\eqref{eq:Qreset_nz}, we obtain the propagator generating function for the BLRW with resetting as
\begin{align}
\widetilde{Q}(n,z|n_0) &= \frac{\eta   }{z q (1-r)  \sinh\Big[   \arccosh  \Big\{  \frac{ \eta - \eta z(1-r)(1-q) }{ z q (1-r)} \Big\}  \Big]} \nonumber \\ 
&\hskip13pt \times  \Bigg[  \frac{r z}{1-z} f^{\frac{n-n_c}{2}} \big\{\alpha\big(z(1-r)\big) \big\}^{-|n-n_c|}  +   f^{\frac{n-n_0}{2}} \big\{\alpha\big(z(1-r)\big) \big\}^{-|n-n_0|} \Bigg] \,  , \label{eq:Qreset_nz_full}
\end{align}
where we have
\begin{align}
\eta \equiv \frac{1}{\sqrt{1-g^2}} \, , \quad f \equiv \frac{1-g}{1+g} \, , \quad
\alpha(z) \equiv \exp \Big[  \arccosh  \Big\{  \frac{\eta}{z q} \big( 1 - z(1-q) \big) \Big\}  \Big]  \, . \label{eq:eta-f-alpha-def}
\end{align}
Note that here $\eta$ and $\alpha$ are independent of the sign of the bias parameter $g$, i.e., the direction of the bias, and one has 
\begin{align}
\sqrt{\eta^2-1} = \eta |g| \, , \quad\quad \eta  +   \sqrt{\eta^2-1}  =  
 \begin{cases} 
          f^{-1/2} & g>0 \, ,\\
          f^{1/2} & g<0 \, . \label{eq:Qreset_MFPT-xx2}
 \end{cases}
\end{align}
Substituting Eq.~\eqref{eq:defect-free-prop} in Eq.~\eqref{eq:Qreset_SS} and using the relation $\sinh[\arccosh(x)] = \sqrt{x^2-1}$, one obtains the stationary-state probability for the resetting dynamics in 1D unbounded domain as
\begin{align}
{}^{\mathrm{ss}}{Q}(n) = \frac{   \eta \, \chi \,  f^{\frac{n-n_c}{2}}   }{\sqrt{\eta^2(1+\chi)^2 -1}}
~ \exp\Big[ - |n-n_c| \arccosh\Big\{ \eta (1+\chi) \Big\} \Big]  \, ,  \label{eq:ss_reset_unbounded}
\end{align}
where the parameter $\chi$ is defined by 
\begin{align}
\chi \equiv \frac{r}{q(1-r)} \, . \label{eq:chi-def}
\end{align}
Note that the parameter $\chi \ge 0$ gives a comparison between resetting and the underlying diffusion, such that large $\chi$ indicates that the BLRW dynamics is dominated by resetting over diffusion, and the opposite for small $\chi$. 
For a given nonzero value of $q$, in the limit $r\to 1 ~(\chi \to \infty)$, i.e., in presence of frequent resetting, from Eq.~\eqref{eq:ss_reset_unbounded} one gets $ {}^{\mathrm{ss}}{Q}(n) \to \delta_{n,n_c}$, as the walker gets stuck at the resetting site $n_c$.

\begin{figure}[htbp]
\centering
\includegraphics[scale=1.25]{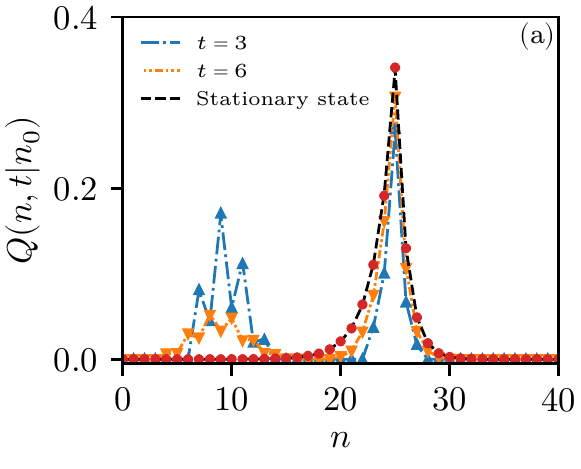} \hspace{10pt}
\includegraphics[scale=1.25]{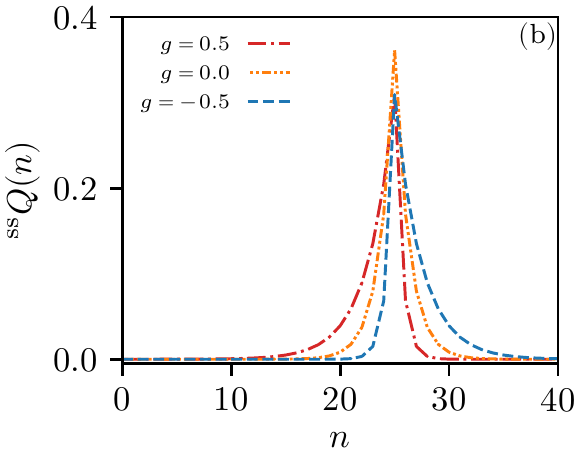} \\[0.2ex]
\includegraphics[scale=1.25]{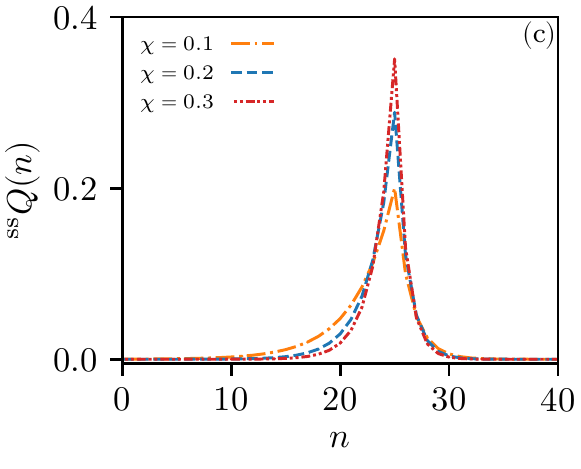}
\caption{\textbf{Propagator and the stationary-state probability for the BLRW with resetting in unbounded domain.}
(a) Time evolution of the propagator ${Q}(n,t|n_0)$ with $r=0.2,~q=0.9,$ and $g=0.2$, i.e., a small bias towards the left. The walker starts from site $n_0=10$. The lines are obtained by numerically inverting~\cite{abate_numerical_1992} Eq.~\eqref{eq:Qreset_nz_full}, while the points are obtained from $10^5$ stochastic  realizations. 
(b) Stationary-state probability ${}^{\mathrm{ss}}{Q}(n)$ from Eq.~\eqref{eq:ss_reset_unbounded} for different values of $g$ with $\chi=0.3$. 
(c) Stationary-state probability ${}^{\mathrm{ss}}{Q}(n)$ for different values of $\chi$ with $g=0.2$. In all three panels the resetting site is at $n_c=25$.}
\label{fig:QReset-unbounded}
\end{figure}
Figure~\ref{fig:QReset-unbounded}\,~shows the time-dependent propagator ${Q}(n,t|n_0)$ and the stationary-state probability ${}^{\mathrm{ss}}{Q}(n)$ for the 1D BLRW with resetting in unbounded domain. Panel (a) shows the relaxation of the propagator to the stationary state. 
As time increases, the peak of the propagator shifts to the resetting site $n_c$.  
Panel (b) depicts the stationary-state probability ${}^{\mathrm{ss}}{Q}(n)$ obtained from Eq.~\eqref{eq:ss_reset_unbounded} for given values of $\chi$ and $n_c$ with varying $g$. From the plot, it is seen that the stationary-state probability is peaked at the resetting site $n_c=25$ and shows a skewness towards the direction of the bias when $g \neq 0$.
For a given value of $g$, the dependence of ${}^{\mathrm{ss}}{Q}(n)$ on the parameter $\chi$ is shown in panel (c). The plot clearly shows that as $\chi$ increases, the width of the stationary-state probability decreases and the probability to find the walker in the neighborhood of the resetting site increases. This is because with an increase in $\chi$ resetting dominates the underlying diffusive dynamics.

For an unbiased lattice walker ($g=0$, $f=\eta=1$) with resetting, the discrete-space stationary-state probability reduces to 
$ {}^{\mathrm{ss}}{Q}(n) = \exp\big[ - |n-n_c| \arccosh (1+\chi) \big] / {\sqrt{1+ 2/ \chi }}  $.
In this case, one may obtain the continuous-space analog of the stationary state by using $x = n L$, $R = r/ T $, and taking the simultaneous limits $n \to \infty, \, L \to 0$ keeping $x$ finite and $r \to 0, \, T \to 0 $ keeping $R$ finite. 
In the limit $r \to 0$, to the leading order in $r$ one has $\chi = r/q$ (see Eq.~\eqref{eq:chi-def}), and expressing the inverse hyperbolic cosine function as a logarithm one has
$\arccosh( 1+ \chi ) = \log \big( 1+ \chi + \sqrt{ 2\chi  + \chi^2  } \, \big)  \simeq \sqrt{2\chi} $, thus yielding
$ {}^{\mathrm{ss}}{Q}(n)  = \sqrt{r/2q} \,  \exp\big(- \sqrt{2 r/q} ~|n-n_c| \big) $. 
In continuous space, the stationary-state distribution ${}^{\mathrm{ss}}{P}(x)$ is then obtained using the relation 
$ {}^{\mathrm{ss}}{P}(x) \Delta x = {}^{\mathrm{ss}}{Q}(n) \Delta x/ L  $
with $R = r/T, \, n=x/L,$ and $n_c=x_c/L$. 
Thus, for the resetting dynamics, we recover the well-known continuous-space stationary-state distribution~\cite{evans_diffusion_2011,evans_diffusion_2011-1} 
\begin{align}
{}^{\mathrm{ss}}{P}(x) = \frac{\alpha_0}{2} ~\ee^{- \alpha_0 |x - x_c|} \,  , \label{eq:ss_reset_unbounded-conti}
\end{align}
where $\alpha_0 \equiv  \sqrt{R/D} $ with the diffusion constant $D \equiv \lim_{L,T \to 0} {q L^2}/{(2T)} $.
Here, the quantity $\alpha_{0}^{-1}$ represents the characteristic diffusion length between resetting events~\cite{evans_diffusion_2011,evans_diffusion_2011-1}.
Note that the parameter $q$ in discrete space appears only as a scaling to the diffusion constant $D$, while $T^{-1}$ gives the number of displacements per unit time.

\subsubsection{First-passage probability}
\label{subsubsec:fpp-unbounded}

For the BLRW with resetting, the generating function of the first-passage probability is obtained from Eqs.~\eqref{eq:defect-free-prop} and~\eqref{eq:Qreset_FP'} as  
\begin{align}
\widetilde{\mathcal  F}(n,z|n_0) = 1 - \frac{(1-z) \Big[ 1- f^{\frac{n-n_0}{2}} \big\{\alpha\big(z(1-r)\big) \big\}^{-|n-n_0|} \Big] }{1-z+ r z f^{\frac{n-n_c}{2}} \big\{\alpha\big(z(1-r)\big) \big\}^{-|n-n_c|} }  \, , \label{eq:Qreset_FP_full}
\end{align} 
which, in the limit $r \to 0$, yields Eq.~\eqref{eq:FP_no_reset_unbounded}.
Similarly, substituting Eq.~\eqref{eq:defect-free-prop} in Eq.~\eqref{eq:Qreset_MFPT'}, one obtains the MFPT to reach $n$ starting from $n_0$ for the BLRW with resetting  as
\begin{align}
T_{n_0 \to n} &= \frac{  1- f^{\frac{n-n_0}{2}} \big[\alpha(1-r) \big]^{-|n-n_0|} }{r f^{\frac{n-n_c}{2}} \big[\alpha(1-r) \big]^{-|n-n_c|} } \, . \label{eq:Qreset_MFPT_full}
\end{align}
For the unbiased walker ($f=\eta=1$) with resetting, multiplying Eq.~\eqref{eq:Qreset_MFPT_full} by the time duration $T$ of a single step and taking the simultaneous limits $r \to 0$ and $T \to 0$ with $R=r/T$ finite; $n_0 \to \infty$ and $L \to 0$ with $x_0=n_0L$ finite; $n_c \to \infty$ and $L \to 0$ with $x_c=n_c L$ finite; and $n \to \infty$ and $L \to 0$ with $x=n L$ finite, one recovers the MFPT in continuous space-time~\cite{evans_diffusion_2011,evans_diffusion_2011-1} as
\begin{align}
T^{\mathrm{conti}}_{x_0 \to x} =   \frac{\ee^{\alpha_0 |x - x_c|}}{R} \,\Big( 1 - \ee^{-\alpha_0 |x - x_0|}  \Big)    \, .\label{eq:Qreset_unbounded_MFPT_conti}
\end{align}

One of the most striking feature of resetting is that it optimizes the mean time to find a target for a diffusive searcher, such that the mean search-time shows a minimum value at an optimal resetting rate~\cite{evans_diffusion_2011,evans_diffusion_2011-1}.
In our case, as the resetting probability $r$ is varied, $T_{n_0 \to n}$ shows a minimum for an optimal value $r=r^{\star}$, which is obtained from the root of the  equation $[\partial T_{n_0 \to n} /\partial r]_{r =r^{\star}} = 0$, namely, 
\begin{align}
r^{\star} |n-n_0|\alpha'(1-r^{\star}) + &  \Big\{ \alpha(1-r^{\star}) + r^{\star} |n-n_c| \alpha'(1-r^{\star}) \Big\} \nonumber \\
&\hskip50pt \times \Big[ f^{-\frac{n-n_0}{2}} \big\{\alpha(1-r^{\star})\big\}^{|n-n_0|} - 1    \Big]  = 0 \, , \label{eq:Qreset_UNB_rOpti}
\end{align}
where we have $\alpha'(z)=\partial \alpha /\partial z$.

In the limit $r\to 0$, using the relation $\exp[\arccosh(z)] = z + \sqrt{z^2-1}$, one obtains from Eq.~\eqref{eq:eta-f-alpha-def}  to the leading order in $r$  that $\alpha(1-r) = ( \eta+\sqrt{\eta^2-1}  ) [1+ \eta r /(q\sqrt{\eta^2 - 1})]$.
Using Eqs.~\eqref{eq:Qreset_MFPT-xx2} and~\eqref{eq:Qreset_MFPT_full}, in the limit $r \to 0$, we then obtain
\begin{align}
T_{n_0 \to n} &= f^{-\frac{n-n_c\pm|n-n_c|}{2}}  \bigg[ \frac{1- f^{\frac{n-n_0\pm|n-n_0|}{2}}}{r}  +  f^{\frac{n-n_0\pm|n-n_0|}{2}} \frac{|n-n_0|}{q|g|} \nonumber \\
&\hskip150pt + \frac{|n-n_c|}{q|g|} \bigg( 1 - f^{\frac{n-n_0\pm|n-n_0|}{2}} \bigg) \bigg] + \mathcal{O}(r) \, , \label{eq:Qreset_MFPT_full-nc0-small-r}
\end{align}
where the $+$ sign corresponds to $g>0$, and the $-$ sign corresponds to $g<0$. Note that as $r $ approaches zero the first term on the right-hand side of Eq.~\eqref{eq:Qreset_MFPT_full-nc0-small-r}, i.e., the leading order term in $r$ shows a $1/r$ divergence. Additionally, $T_{n_0 \to n}$ also diverges for a dynamics without any bias, i.e., with $g=0$.
However, for $g>0$ and $n < n_0$ as well as for $g<0$ and $n > n_0$, i.e., for the case when the target is located in the direction of the bias, the $1/r$ diverging term vanishes, and  $T_{n_0 \to n}$ approaches a saturation value $T^{\mathrm{sat}}_{n_0 \to n}$ given by
\begin{align}
T^{\mathrm{sat}}_{n_0 \to n} =  \frac{  |n - n_0|     }{q|g|} ~ f^{-\frac{n-n_c\pm|n-n_c|}{2}} \, , \label{eq:Qreset_MFPT_full-nc0-small-r-sat}
\end{align}
where the $\pm$ sign as before changes with the sign of $g$. Despite the fact that Eq.~\eqref{eq:Qreset_MFPT_full-nc0-small-r-sat} is obtained in the limit $r \to 0$, the saturation value $T^{\mathrm{sat}}_{n_0 \to n}$ does not reduce to the reset-free case given in Eq.~\eqref{eq:QNR_MFPT} (except when $r=0$), it rather bears the signature of the resetting site $n_c$. This is because no matter how negligible $r$ might be, even a single resetting event would affect the dynamics.

\begin{figure}[htbp]
\centering
\includegraphics[scale=1.1]{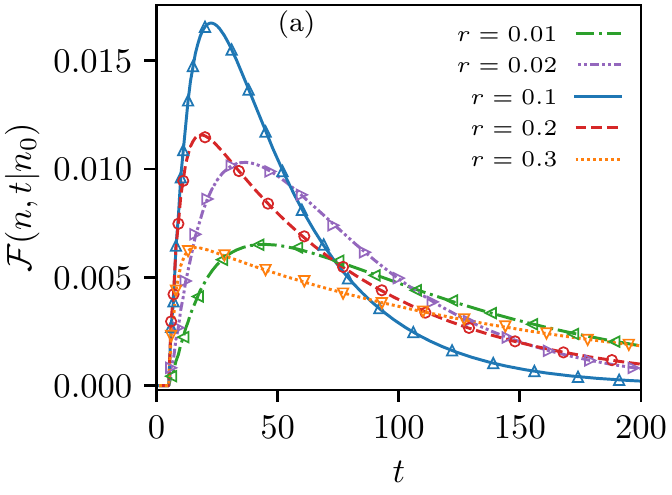} \hspace{5pt}
\includegraphics[scale=1.1]{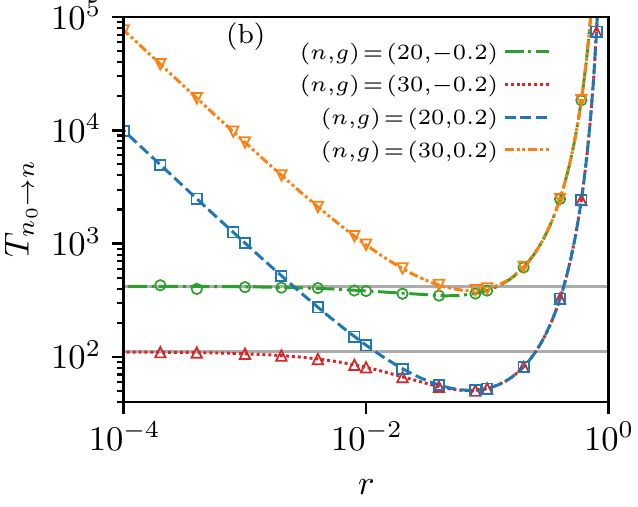}
\caption{\textbf{First-passage probability and the MFPT for the BLRW with resetting in unbounded domain.}
(a) First-passage probability ${\mathcal  F}(n,t|n_0)$ for different values of resetting probability $r$ with $q=0.9, \,g=0.2$. The walker starts from site $n_0=10$, its motion is subject to resetting to site $n_c=25$, and the target site is at $n=20$. 
The lines are obtained by numerically inverting~\cite{abate_numerical_1992} Eq.~\eqref{eq:Qreset_FP_full}, while the points are obtained from $5\times 10^6$ stochastic realizations. 
(b) MFPT $T_{n_0 \to n} $ as a function of $r$ with $n_0=10,\, n_c=25,$ and $ q=0.9$. The broken lines are obtained from Eq.~\eqref{eq:Qreset_MFPT_full}, whereas the points are obtained from $10^5$ stochastic realizations. The horizontal solid lines show the saturation value $T^{\mathrm{sat}}_{n_0 \to n}$ obtained from Eq.~\eqref{eq:Qreset_MFPT_full-nc0-small-r-sat}. }
\label{fig:QReset-unbounded-FP}
\end{figure}
Figure~\ref{fig:QReset-unbounded-FP}(a) depicts the non-monotonic behavior of the first-passage probability as a function of $r$.  
While with an increase in $r$ the mode of the first-passage probability decreases, the corresponding mode probability changes non-monotonically, i.e., it first increases and then decreases, suggesting the occurrence of extrema for the MFPT as $r$ is varied. 
Figure~\ref{fig:QReset-unbounded-FP}(b) shows the appearance of such extrema for different values of $n$ and $g$. 
The optimal value $r=r^{\star}$ at which $T_{n_0 \to n}$ has a minimum is given by the root of Eq.~\eqref{eq:Qreset_UNB_rOpti}. As we have $n>n_0$ in the plot, with decreasing $r$ the quantity $T_{n_0 \to n}$ saturates to the value $T^{\mathrm{sat}}_{n_0 \to n}$ for $g<0$ and diverges for $g>0$.

\begin{figure}[htbp]
\centering
\includegraphics[scale=1.1]{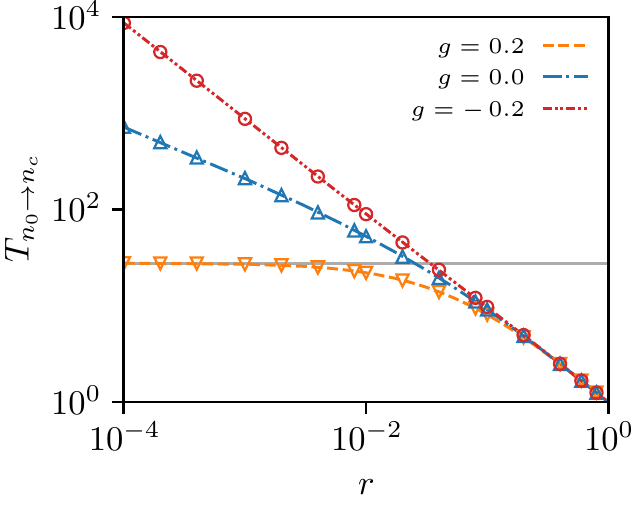} 
\caption{\textbf{MFPT to the resetting site $n_c$ for the BLRW in unbounded domain $T_{n_0 \to n_c} $ as a function of $r$ for different values of $g$ with $n_0=30,\,n_c=25,$ and $q=0.9$.} The broken lines are obtained by using Eq.~\eqref{eq:Qreset_MFPT_full-nc0}, while the points are obtained from $10^5$ stochastic realizations. The horizontal line shows the saturation value $T^{\mathrm{sat}}_{n_0 \to n_c} $ obtained from Eq.~\eqref{eq:Qreset_MFPT_full-nc0-small-r-sat}. }
\label{fig:QReset-unbounded-MFPT-at-NC}
\end{figure}

The MFPT to the resetting site $n_c$ starting from site $n_0$ obtained from Eq.~\eqref{eq:Qreset_MFPT_full} is given by
\begin{align}
T_{n_0 \to n_c} =  \frac{1}{r}-  \frac{f^{\frac{n_c-n_0}{2}}}{r}~ \exp\Bigg[-|n_c-n_0|\arccosh  \Bigg\{  \frac{ \eta - \eta (1-r)(1-q) }{(1-r)q} \Bigg\}\Bigg]  \, ,  \label{eq:Qreset_MFPT_full-nc0}
\end{align}
where we have used Eq.~\eqref{eq:eta-f-alpha-def}. For $n_0 \neq n_c$, in the limit $r \to 1$, the second term on the right-hand side of Eq.~\eqref{eq:Qreset_MFPT_full-nc0} approaches zero because of the exponential factor, and $T_{n_0 \to n_c}$ becomes independent of the bias. As $T_{n_0 \to n_c} = 0$ when $n_0 = n_c$, with resetting occuring at every time step, i.e., in the limit $r \to 1$, one gets $T_{n_0 \to n_c} \to (1-\delta_{n_0,n_c})/r$.
Figure~\ref{fig:QReset-unbounded-MFPT-at-NC}\, shows the MFPT to the resetting site as a function of $r$ for different values of $g$.
In the limit $r \to 0$, the quantity $T_{n_0 \to n_c}$ diverges as usual for the unbiased case, and also for the biased case when the target $n_c$ is located opposite to the direction of the bias, but when $n_c$ is located in the direction of the bias,  $T_{n_0 \to n_c}$ approaches the saturation value $T^{\mathrm{sat}}_{n_0 \to n_c} $. It is also seen that in the limit $r \to 1$,   $T_{n_0 \to n_c}$ becomes independent of $g$ and indeed behaves as $1/r$. 
The mean return time to site $n$ with resetting obtained from Eqs.\eqref{eq:Qreset_MRT'} and~\eqref{eq:ss_reset_unbounded} is given by 
$ R(n) = \sqrt{(1+1/\chi)^2 -1/(\eta\chi)^2}  \, f^{\frac{n_c-n}{2}} \, \exp\big[ |n-n_c| \arccosh\{ \eta (1+\chi) \} \big]  $,
which in the absence of bias yields $R(n) = \sqrt{1+2/\chi} \exp\big[ |n-n_c| \arccosh (1+\chi) \big]$.

\subsubsection{Partially absorbing trap}
\label{subsubsec:part-abs-trap-unbounded}

In the presence of a partially absorbing trap at site $n_{\mathrm d}$, the survival probability $S(t|n_0)$ and the absorption probability $\mathcal{A}(n_{\mathrm d},t|n_0)$ for the resetting dynamics are obtained by substituting Eq.~\eqref{eq:defect-free-prop} in Eq.~\eqref{eq:Qreset_SP_in_reset_free_SP} and Eq.~\eqref{eq:Qreset_AP'}, respectively, and then employing a $z$-inversion.
Similarly, in this case the mean absorption time obtained from Eq.\eqref{eq:Qreset_MFAT'} is given by
\begin{align}
A_{n_0 \to n_{\mathrm d}} =  f^{\frac{n_c-n_{\mathrm d}}{2}} \big[ \alpha(1-r)\big]^{|n_{\mathrm d}-n_c|} & \Bigg[ \frac{1- f^{\frac{n_{\mathrm d}-n_0}{2}} \big[ \alpha(1-r)\big]^{-|n_{\mathrm d}-n_0|}  }{r} \nonumber \\ 
&\hskip30pt + \left( \frac{1}{\rho} -1 \right) \frac{\sqrt{\eta^2(1+\chi^2)-1}}{\eta \chi} \Bigg]  \, .\label{eq:Qreset_unbounded_MFAT}
\end{align}
For the unbiased walker ($f=\eta=1$) with resetting, multiplying Eq.~\eqref{eq:Qreset_unbounded_MFAT} by the time duration $T$ of a single step and once again taking the continuous limits (as described for Eq.~\eqref{eq:Qreset_unbounded_MFPT_conti}), one obtains the mean absorption time at $x_{\mathrm{d}}$ in continuous space-time as
\begin{align}
A^{\mathrm{conti}}_{x_0 \to x_{\mathrm d}} =   \frac{\ee^{\alpha_0 |x_{\mathrm d} - x_c|}}{R} \,\Big( 1 - \ee^{-\alpha_0 |x_{\mathrm d} - x_0|}  \Big) + \frac{2 \, \ee^{\alpha_0 |x_{\mathrm d} - x_c|} }{c \, \alpha_0} \, , \label{eq:Qreset_unbounded_MFAT_g=0_conti}
\end{align}
where the constant $c$ is the absorption velocity defined by $c \equiv \lim_{L,T,\rho\to 0} \, L \rho/\big(T(1-\rho)\big) $.
Equation~\eqref{eq:Qreset_unbounded_MFAT_g=0_conti} reduces to the known result for the mean absorption time in diffusion with resetting~\cite{whitehouse_effect_2013}, namely, $A^{\mathrm{conti}}_{x_0 \to 0} = [ \exp(\alpha_0 |x_0|) -1 ]/R + 2\, \exp(\alpha_0 |x_0|)  / (c \, \alpha_0)   $ when $x_c = x_0$ and $x_{\mathrm d}=0$.

\subsection{Bounded domain}
\label{subsec:prop-bounded}

We consider the same resetting dynamics of the BLRW, but now the walker is (i) on a periodic domain or (ii) limited by two reflecting boundaries at sites $n=1$ and $n=N$. The underlying diffusive dynamics at the boundaries is now modified, while that in the bulk remains the same. 
In periodic domain, from $n=1$ the walker moves to $n=N$ with probability $q(1+g)/2$, and from $n=N$ to $n=1$ with probability $q(1-g)/2$. 
For reflecting boundaries, from $n=1$ the walker moves to the right with probability $q(1-g)/2$ or stays with probability $(1-q(1-g)/2)$, and similarly from $n=N$ it moves to the left with probability $q(1+g)/2$ or stays with probability  $(1-q(1+g)/2)$. 
The propagators for the reset-free BLRW dynamics in 1D bounded domain are taken from Ref.~\cite{sarvaharman_closed-form_2020} and reported in~\ref{sec:BLRW}.

\subsubsection{Occupation probability}
\label{subsubsec:occu-prop-bounded}

We apply the same framework developed in Sec.~\ref{sec:TheResettingModel}\, and obtain the time-dependent bounded propagator 
\begin{align}
{Q}({n},t|n_{{0}} )  &=  r \sum_{k = 0}^{N - 1}  h^{(N)}_{k} (n, n_{{c}})  ~\frac{\big({\gamma^{(N)}_{k}}\big)^{t}- 1}{{\gamma^{(N)}_{k}}-1} +   \sum_{k = 0}^{N - 1} h^{(N)}_{k} (n, n_{{0}})  ~\big({\gamma^{(N)}_{k}}\big)^{t} \, , \label{eq:QReset-bounded_time_dep_prop} 
\end{align}
where one has
\begin{align}
\gamma^{(N)}_{k} \equiv (1-r) \Big[ 1+ s^{(N)}_{k} \Big]  \, . \label{eq:gammaki-def-noi}
\end{align}
Equation~\eqref{eq:QReset-bounded_time_dep_prop} is obtained by inserting Eq.~\eqref{eq:PsiNR-gen} in Eq.~\eqref{eq:reset-x},
and performing the time summation. 
The quantities $h^{(N)}_{k} (n, m)$ and $s^{(N)}_{k}$, respectively, the eigenvectors and the eigenvalues of the lattice walk transition matrix in the master equation, are given in Eqs.~\eqref{eq:PsiNR-peri-h},~\eqref{eq:PsiNR-peri-s} for periodic boundaries and in Eqs.~\eqref{eq:PsiNR-refl-h},~\eqref{eq:PsiNR-refl-s} for reflecting boundaries.
The propagator generating function $\widetilde{Q}(n,z|n_0)$ is obtained by substituting Eq.~\eqref{eq:QNR_bounded} in Eq.~\eqref{eq:Qreset_nz}, while the stationary-state probability, obtained from Eqs.~\eqref{eq:QNR_bounded} and~\eqref{eq:Qreset_SS}, is given by
\begin{align}
{}^{\mathrm{ss}}{Q}(n) = \sum_{k = 0}^{N - 1} \frac{ \chi \,h^{(N)}_{k} (n, n_c)}{\chi-  \overline{s}^{(N)}_{k} } \, , \label{eq:Qreset_bounded_SS}
\end{align}
whose limit for $r \to 1 ~(\chi \to \infty)$ is ${}^{\mathrm{ss}}{Q}(n) = \sum_{k = 0}^{N - 1} h^{(N)}_{k} (n, n_c) = \delta_{n,n_c} $.
In Eq.~\eqref{eq:Qreset_bounded_SS}, we have introduced the $q$-independent quantity $ \overline{s}^{(N)}_{k} \equiv {s}^{(N)}_{k} /q $ (see Eqs.~\eqref{eq:PsiNR-peri-s} and~\eqref{eq:PsiNR-refl-s}).

\begin{figure}[!htbp]
\centering
\includegraphics[scale=1.2]{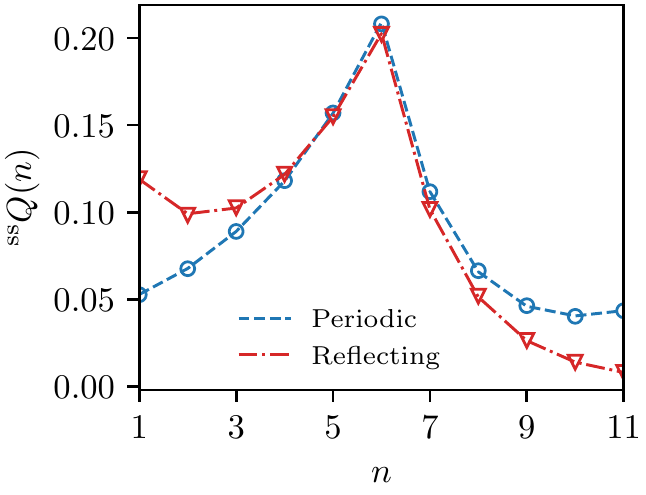}
\caption{{\textbf{Stationary-state probability for the BLRW with resetting ${}^{\mathrm{ss}}{Q}(n)$ in the finite domain $[1,N]$ with $N=11, \, n_c = 6, \, \chi = 0.1, $ and $g=0.2$.}} The lines are obtained by using Eq.~\eqref{eq:Qreset_bounded_SS}, while the points are obtained from $10^6$ stochastic realizations. The dashed and dash-dotted lines show, respectively, the periodic and reflecting cases.}
\label{fig:QReset_bounded_SS}
\end{figure}
Figure~\ref{fig:QReset_bounded_SS}\, depicts the stationary-state probability ${}^{\mathrm{ss}}{Q}(n)$ for the BLRW with resetting in bounded domain. The quantity ${}^{\mathrm{ss}}{Q}(n)$ is peaked at the resetting site $n_c$ for both the periodic and reflecting boundaries. As the bias $g$ is positive, ${}^{\mathrm{ss}}{Q}(n)$ shows a skewness towards the negative $x$-axis in both cases. For the periodic case because of the periodicity of the lattice there is not much difference between the stationary values of the probability at the boundaries. On the other hand, for the reflecting case, the walker in the stationary state is much more likely to be found at the boundary in the direction of the bias rather than that against it. Note that in reflecting domain for strong enough bias the stationary probability at the boundary in the direction of the bias can be higher than that at ${n}_c$.

\subsubsection{First-passage probability}
\label{subsubsec:fpp-bounded}

The generating function of the first-passage probability ${\mathcal F}(n,t|n_0)$ in this case is obtained by substituting Eq.~\eqref{eq:QNR_bounded} in Eq.~\eqref{eq:Qreset_FP'} as
\begin{align}
\widetilde{\mathcal  F}(n,z|n_0) &= \Bigg[ \sum_{k = 0}^{N - 1} \frac{r z h^{(N)}_{k} (n, n_c)+ (1-z) h^{(N)}_{k} (n, n_0) }{1-z (1-r)  \Big( 1+ s^{(N)}_{k} \Big)}   \Bigg] \nonumber \\
&\hskip50pt \times \Bigg[ \sum_{k = 0}^{N - 1} \frac{r z h^{(N)}_{k} (n, n_c)+ (1-z) h^{(N)}_{k} (n, n) }{1-z (1-r)  \Big( 1+ s^{(N)}_{k} \Big)}  \Bigg]^{-1} \, . \label{eq:QReset_bounded_FPT}
\end{align}
Similarly, we get the MFPT to site $n$ starting from site $n_0$ from Eqs.~\eqref{eq:QNR_bounded} and~\eqref{eq:Qreset_MFPT'} as 
\begin{align}
T_{n_0 \to n} = \Bigg[\sum_{k=0}^{N-1} \frac{  h^{(N)}_{k} (n, n) - h^{(N)}_{k} (n, n_0) }{r-(1-r)  s^{(N)}_{k} } \bigg] \times \Bigg[   \sum_{k=0}^{N-1} \frac{ r \,h^{(N)}_{k} (n, n_c)}{r-(1-r)  s^{(N)}_{k} }\Bigg]^{-1}  \, .  \label{eq:Qreset_bounded_MFT}
\end{align}
In the limit $r \to 0$, unlike the $1/r$ divergence in the unbounded domain (see Eq.~\eqref{eq:Qreset_MFPT_full-nc0-small-r}), here the MFPT $T_{n_0 \to n}$ has an $r$-independent leading-order behavior given by
\begin{align}
T^{\mathrm{sat}}_{n_0 \to n} = \frac{1}{h^{(N)}_{0}(n,n_c)} \sum_{k=1}^{N-1} \frac{h^{(N)}_{k} (n, n_0) - h^{(N)}_{k} (n, n)}{s^{(N)}_{k}} \, . \label{eq:Qreset_bounded_MFT_sat}
\end{align}
As the quantity $h^{(N)}_{0}(n,n_c)$ is independent of $n_c$ for both periodic and reflecting domains, the saturation value of the MFPT in bounded domain does not bear any signature of resetting. The MFPT obtained in the limit $r\to 0$ and given in Eq.~\eqref{eq:Qreset_bounded_MFT_sat} is indeed the MFPT for the reset-free BLRW dynamics reported in Ref.~\cite{sarvaharman_closed-form_2020}.

\begin{figure}[htbp]
\centering
\includegraphics[scale=1.125]{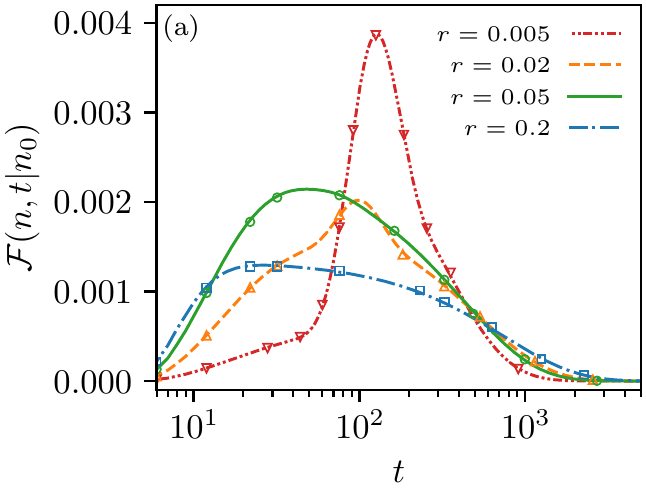} \hspace{5pt}
\includegraphics[scale=1.125]{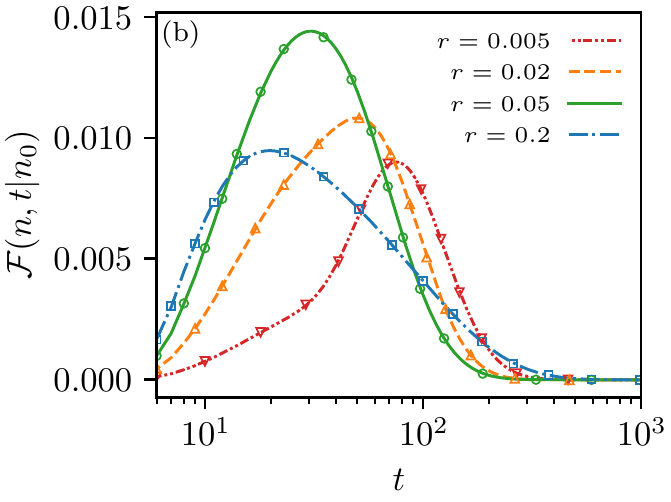}
\caption{\textbf{Time dependence of the first-passage probability ${\mathcal  F}(n,t|n_0)$ for the BLRW with resetting in periodic domain.} The walker starting from site $n_0=10$ moves with $q=0.8$ in a domain of size $N=49$. Its motion is subject to resetting to site $n_c=25$ with different resetting probabilities $r$, while the target site is at $n=30$. (a) ${\mathcal  F}(n,t|n_0)$ for positive bias with $g=0.2$. (b) ${\mathcal  F}(n,t|n_0)$ for negative bias with $g=-\,0.2$. The lines are obtained by numerically inverting~\cite{abate_numerical_1992} Eq.~\eqref{eq:QReset_bounded_FPT}, while the points are obtained from $5\times 10^6$ stochastic realizations.}
\label{fig:Qreset-bounded-FPT-n30}
\end{figure}
In the periodic domain, a target can be reached from two opposite directions. For a reset-free BLRW in periodic domain, one can have bimodal first-passage probability ${\mathcal  F}(n,t|n_0)$, when the shortest direction to the target is opposite to the bias~\cite{sarvaharman_closed-form_2020}. 
Figure~\ref{fig:Qreset-bounded-FPT-n30}\, shows how resetting modifies this first-passage probability and adds rich features to it.   
For panel (a), the bias is opposite to the shortest direction, and the resetting site lies on the shortest path. As a result, for small non-zero values of $r$, resetting and the bias act oppositely in reaching the target: while resetting reduces the distance to the target, the bias forces the walker to move along the longest path. Because of these two contrasting influences, the first-passage probability shows flatter peaks (see the lines for $r=0.02$ and $0.05$ in panel (a)).
For panel (b), the bias is along the shortest direction, and the resetting site lies on the shortest path. Consequently, for small non-zero values of $r$, both resetting and the bias effectively aid in reaching the target and thus make a quicker first-passage more probable.
However, for large enough values of $r$, resetting dominates the dynamics over the bias making the first-passage less probable, which leads to the fat tails of ${\mathcal  F}(n,t|n_0)$ (see the line for $r=0.2$ in both panels (a) and (b)).  
For both the panels, as $r$ increases, the mode of the first-passage probability decreases, but the mode probability shows non-monotonic behavior. In panel (a), due to the competing nature of resetting and the bias, the mode probability shows two-fold non-monotonicity: with increasing $r$ it first decreases (from $r=0.005$ to $r=0.02$), then increases (from $r=0.02$ to $r=0.05$), and again decreases (from $r=0.05$ to $r=0.2$). Similar two-fold non-monotonic behavior of ${\mathcal F}(n,t|n_0)$ is also detectable at the tails of the probability for large values of $t$. 
On the other hand, in panel (b), the non-monotonicity of the mode probability is single-fold, namely, as $r$ increases, it first increases (from $r=0.005$ to $r=0.05$) and then decreases (from $r=0.05$ to $r=0.2$). This single-fold non-monotonicity is also seen at the tails of the probability ${\mathcal  F}(n,t|n_0)$.

\begin{figure}[htbp]
\centering
\includegraphics[scale=1.125]{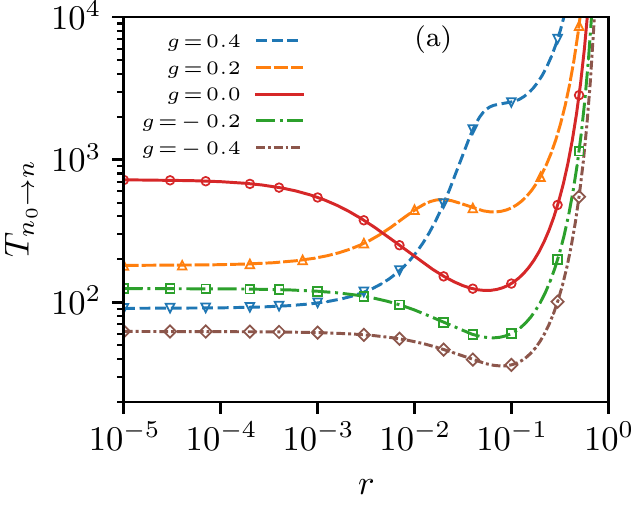} \hspace{5pt}
\includegraphics[scale=1.125]{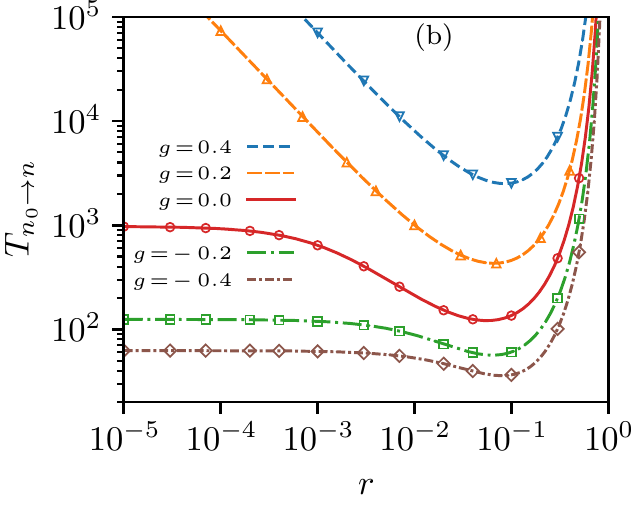}
\caption{\textbf{MFPT $T_{n_0 \to n}$ for the BLRW in bounded domains as a function of $r$ for different biases.} 
Other parameters are the same as in fig.~\ref{fig:Qreset-bounded-FPT-n30}.
(a) Periodic domain. (b) Reflecting domain. The lines are obtained by using Eq.~\eqref{eq:Qreset_bounded_MFT}, while the points are obtained from $10^5$ stochastic realizations. For all the plots, the MFPT, in the limit $r \to 0$, approaches the saturation value $T^{\mathrm{sat}}_{n_0 \to n} $ given in Eq.~\eqref{eq:Qreset_bounded_MFT_sat}.}
\label{fig:Qreset-bounded-MFPT-n30}
\end{figure}
Figure~\ref{fig:Qreset-bounded-MFPT-n30}\, shows the MFPT $T_{n_0 \to n}$ as a function of $r$ for different values of $g$ in the bounded domain. 
Panels (a) and (b) depict the results for the periodic and reflecting cases, respectively.
In panel (a), as $r$ is varied, $T_{n_0 \to n}$ shows a minimum value for $g \leq 0$. 
However, for $g=0.2$, the two-fold non-monotonicity of ${\mathcal  F}(n,t|n_0)$ observed in Fig.~\ref{fig:Qreset-bounded-FPT-n30}(a) manifests itself in the appearance of more than one extremum value in $T_{n_0 \to n}$ with varying $r$.
With reflecting boundaries in panel (b), $T_{n_0 \to n}$ shows a single minimum as $r$ is varied for all values of $g$. 
\begin{figure}[htbp]
\centering
\includegraphics[scale=1.125]{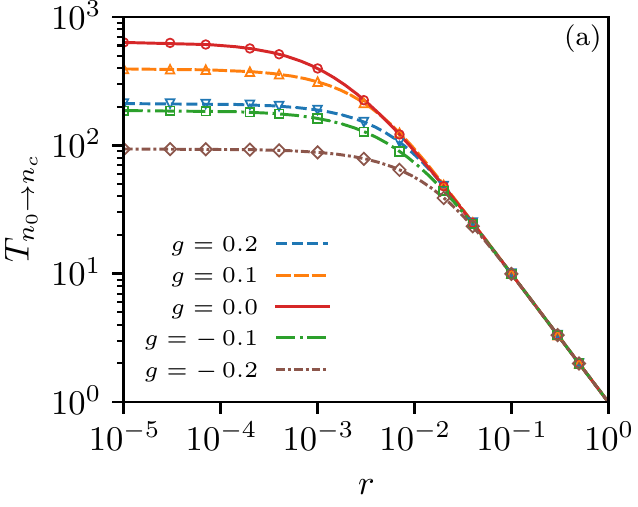} \hspace{5pt}
\includegraphics[scale=1.125]{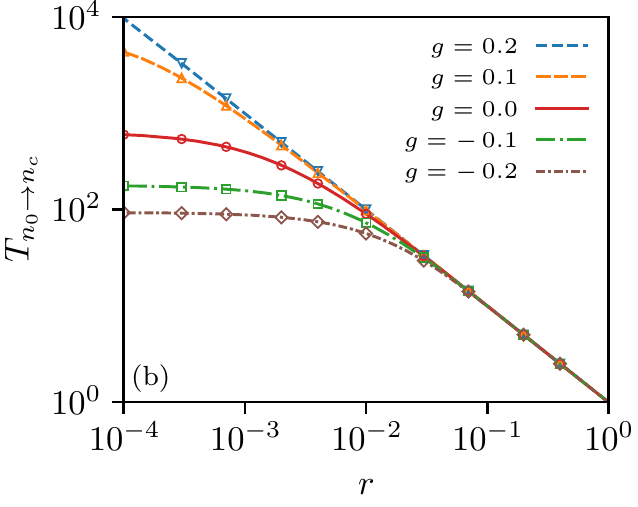}
\caption{\textbf{MFPT to the resetting site $T_{n_0 \to n_c}$ as a function of $r$ for the BLRW in bounded domains.} Parameters are the same as in Fig.~\ref{fig:Qreset-bounded-MFPT-n30}, but with $n = n_c =25$.
(a) Periodic domain. (b)  Reflecting domain. 
The lines are obtained by using Eq.~\eqref{eq:Qreset_bounded_MFT}, while the points are obtained from $10^5$ stochastic realizations.}
\label{fig:Qreset-bounded-MFPT-nc}
\end{figure}
Figure~\ref{fig:Qreset-bounded-MFPT-nc}\, shows the MFPT to the resetting site $n_c$ as a function of $r$ for the BLRW in the bounded domain for different values of $g$. The plots clearly indicate that in the limit $r \to 1$, the quantity $T_{n_0 \to n_c}$ becomes independent of the bias and behaves as $1/r$ similar to the unbounded case shown in Fig.~\ref{fig:QReset-unbounded-MFPT-at-NC}.

\section{Transmission in one dimension}
\label{sec:trans}

We now consider the dynamics of two interacting BLRWs with resetting in 1D bounded domain. The interaction in this case is defined as transmission, such that the walkers when on the same site may with a given probability exchange information or one individual may pass information to the other.
These transmission events may in general be of different types, e.g., transmission of an infectious pathogen from an infected individual to a susceptible one, the capture of a prey by a predator when information transfer implies the death of either of the individuals, or anihilation when both walkers die upon successful transmission.

We denote the positions of the first and the second 1D walker, by $n_1 \in [1,N_1]$ and $n_2 \in [1,N_2]$, respectively, with $N_1$ and $N_2$ being their corresponding lattice sizes.
The resetting sites for the first and the second walker are denoted by $n_{c_1}$ and $n_{c_2}$, respectively.
In a 2D domain of size $N_1 \times N_2$, the position of the combined walker is specified by the vector $\bm{n}=(n_1,n_2)$.
Depending on the nature of the underlying resetting process, the combined dynamics may be classified into two schemes: (i) simultaneous resetting and (ii) independent resetting.
In simultaneous resetting scheme, both of the 1D walkers, despite having independent underlying diffusive dynamics, at any time instant get reset simultaneously to their corresponding resetting sites with probability $r$. As a result, the combined 2D walker at any time instant resets to site $\bm{n}_c=(n_{c_1},n_{c_2})$ with the same probability.
On the other hand, in the independent resetting scheme, both the resetting and the underlying diffusive dynamics of the 1D walkers are independent. In this scheme, the first and the second 1D walker get reset to their corresponding resetting sites with probabilities $r_1$ and $r_2$, respectively.  
Consequently, the 2D walker at any time instant either resets to a site from the set $\mathfrak{C}\equiv\{(n_{c_1},n') | n'\in [1,N_2]\}$ with probability $r_1$; or to a site from the set $\mathfrak{R} \equiv \{(n',n_{c_2})|n'\in [1,N_1]\}$ with probability $r_2$.
For brevity, we use the following vector notations $\bm{r} = (r_1,r_2), \, \bm{q} = (q_1,q_2),$ and $\bm{g} = (g_1,g_2)$, where $r_i,\, q_i,$ and $g_i$ are, respectively, the resetting probability, diffusion parameter, and bias parameter for the $i$-th 1D walker. 

\subsection{Propagator with simultaneous resetting}
\label{subsec:simul-reset}

In this scheme, the 2D propagator is obtained by considering the fact that only the diffusive dynamics of the 1D walkers are independent. In the absence of resetting, the time-dependent propagator for the 2D walker to be at position $\bm{n}$ at time $t$ starting from position $\bm{n_0} = (n_{0_1}, n_{0_2})$ may be written as
\begin{align}
Q_{\mathrm{NR}}(\bm{n}, t | \bm{n_0}) = Q_{\mathrm{NR}}(n_1, t | n_{0_1}) \, Q_{\mathrm{NR}}(n_2, t | n_{0_2}) \, , \label{eq:QNR_2d_simul_gen_in_t-0}
\end{align}
where $Q_{\mathrm{NR}}(n_1, t | n_{0_1})$ and $ Q_{\mathrm{NR}}(n_2, t | n_{0_2})$ are the reset-free time-dependent  propagators for the first and the second 1D walker, respectively. 
Note that Eq.~\eqref{eq:QNR_2d_simul_gen_in_t-0} gives the reset-free time-dependent 2D propagator with next-nearest-neighbor hopping, i.e., the walker from site $\bm{n} = (n_1,n_2)$ at the next instant may go to nine possible sites given by $\bm{n} = (n_1 + i, n_2 +j)$ with $i,j \in [-1,0,1]$.  

Using the expression for the 1D bounded propagator given in Eq.~\eqref{eq:PsiNR-gen}, we obtain the reset-free 2D bounded propagator from Eq.~\eqref{eq:QNR_2d_simul_gen_in_t-0} as
\begin{align}
{{Q}}_{\mathrm{NR}}(\bm{n},t|\bm{n_0}) &= \sum_{k_1 = 0}^{N_1 - 1}   \sum_{k_2 = 0}^{N_2 - 1} ~  h^{(N_1)}_{k_1} (n_1, n_{0_1}) ~  h^{(N_2)}_{k_2} (n_2, n_{0_2}) ~\Big[ 1+ s^{(N_1)}_{k_1} \Big]^t ~\Big[ 1+ s^{(N_2)}_{k_2} \Big]^t \, , \label{eq:QNR_2d_simul_gen_in_t}
\end{align}
for which the generating function is given by
\begin{align}
{\widetilde{Q}}_{\mathrm{NR}}(\bm{n},z|\bm{n_0}) = \sum_{k_1 = 0}^{N_1 - 1}   \sum_{k_2 = 0}^{N_2 - 1} ~ \frac{ h^{(N_1)}_{k_1} (n_1, n_{0_1}) ~  h^{(N_2)}_{k_2} (n_2, n_{0_2}) }{1-z \Big[ 1+ s^{(N_1)}_{k_1} \Big] ~\Big[ 1+ s^{(N_2)}_{k_2} \Big] } \, .  \label{eq:QNR_2d_simul_gen_in_z}
\end{align}
As the 2D walker in this scheme gets reset to site $\bm{n_c}$ with probability $r$, one substitutes Eq.~\eqref{eq:QNR_2d_simul_gen_in_z} in Eq.~\eqref{eq:Qreset_nz} and obtains the generating function of the 2D propagator with resetting as
\begin{align}
{\widetilde{Q}}(\bm{n},z|\bm{n_0}) &= \frac{r z}{1-z}\sum_{k_1 = 0}^{N_1 - 1}   \sum_{k_2 = 0}^{N_2 - 1} ~ \frac{ h^{(N_1)}_{k_1} (n_1, n_{c_1}) ~  h^{(N_2)}_{k_2} (n_2, n_{c_2}) }{1-z(1-r) \Big[ 1+ s^{(N_1)}_{k_1} \Big] ~\Big[ 1+ s^{(N_2)}_{k_2} \Big] } \nonumber \\[1ex]
&\hskip60pt + \sum_{k_1 = 0}^{N_1 - 1}   \sum_{k_2 = 0}^{N_2 - 1} ~ \frac{ h^{(N_1)}_{k_1} (n_1, n_{0_1}) ~  h^{(N_2)}_{k_2} (n_2, n_{0_2}) }{1-z(1-r) \Big[ 1+ s^{(N_1)}_{k_1} \Big] ~\Big[ 1+ s^{(N_2)}_{k_2} \Big] } \, .  \label{eq:Qreset_2d_simul_gen_in_z}
\end{align} 

\begin{figure}[!htbp]
\centering
\includegraphics[scale=0.95]{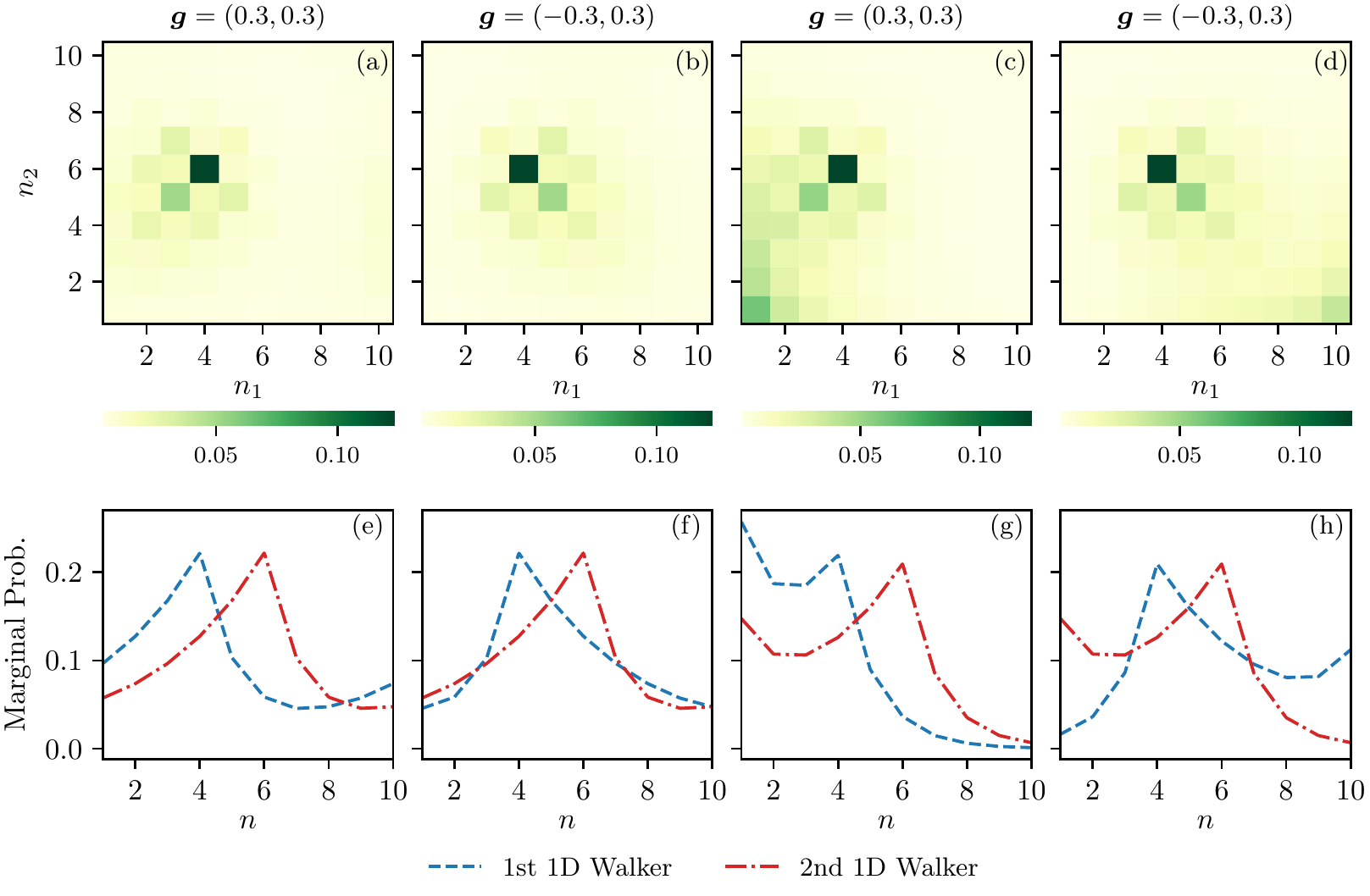} 
\caption{\textbf{2D propagator ${Q}(\bm{n},t|\bm{n_0})$ with simultaneous resetting for different biases at time $t=10^2$.} The plots are obtained by numerically inverting Eq.~\eqref{eq:Qreset_2d_simul_gen_in_z}. The parameters of the system are $N_1=N_2=10$, $\bm{n_c} = (4,6)$, $\bm{n_0} = (4,6)$, $r=0.1$, and $\bm{q}=(0.9,0.9)$. Top row: (a) and (b) ${Q}(\bm{n},t|\bm{n_0})$ in periodic domain, (c) and (d) ${Q}(\bm{n},t|\bm{n_0})$ with reflecting boundaries. Bottom row: Each panel shows the marginal probabilities of the 1D walkers corresponding to the 2D propagator ${Q}(\bm{n},t|\bm{n_0})$ given in the adjacent panel above.   
}
\label{fig:QReset-simul-joint-2d}
\end{figure}
The top row of Fig.~\ref{fig:QReset-simul-joint-2d}\, shows the 2D propagator ${{Q}}(\bm{n},t|\bm{n_0})$ with simultaneous resetting in bounded domain.  
From all four top panels, we see that the probability ${{Q}}(\bm{n},t|\bm{n_0})$ is maximum at the resetting site $\bm{n_c}$. However, note that in reflecting domain for strong enough bias the probability at the boundary in the direction of the bias can be higher than that at $\bm{n}_c$. We recall that for the 1D walkers a positive value of $g$ corresponds to bias towards the negative $x$-axis. As a result, with $g_1>0$ and $g_2>0$ the propagator ${{Q}}(\bm{n},t|\bm{n_0})$ is seen skewed towards the lower-left direction of the 2D domain, while for $g_1<0$ and $g_2 > 0$ the skewness shifts towards the lower-right direction. 
Panels in the bottom row of Fig.~\ref{fig:QReset-simul-joint-2d}\, represent the marginal site-occupation probability of the 1D walkers corresponding to ${Q}(\bm{n},t|\bm{n_0})$ shown in the adjacent corresponding panels above. The marginal site-occupation probability of the $i$-th 1D walker, i.e., the probability to find the walker at site $n_i$ at time $t$ starting from site $n_{0_i}$ when the other 1D walker can be anywhere, is given by $\mathbb{P}(n_i,t|n_{0_i}) =  \sum_{n_i = 1}^{N_i} {Q}(\bm{n},t|\bm{n_0}) $. From panels (e), (f), and (h) one sees that the marginal probabilities of the individual 1D walkers are peaked at their corresponding resetting sites, while in panel (g) the bias for the first 1D walker is strong enough to make it peaked at the reflecting boundary in the direction of the bias. 

\subsection{Propagator with independent resetting}
\label{subsec:indep-reset}

In this scheme, since both the resetting and the diffusive dynamics of the 1D walkers are independent, the time-dependent 2D propagator with resetting is given by
\begin{align}
Q(\bm{n},t|{\bm{n_0}}) &= Q({n_1},t|n_{{0_1}}) ~Q({n_2},t|n_{{0_2}}) \, , \label{eq:QReset-indep-2d-in-t-00}
\end{align}
where $Q({n_i},t|n_{{0_i}})$ is the propagator of the $i$-th 1D BLRW with resetting. The quantities $Q({n_i},t|n_{{0_i}})$ are obtained from Eq.~\eqref{eq:QReset-bounded_time_dep_prop} by replacing the parameters of the set $\{r,n,n_{0},n_{c},N\}$ with the corresponding ones of the set $\{r_i,n_i,n_{0_i},n_{c_i},N_i\}$.
The explicit expression for $Q(\bm{n},t|{\bm{n_0}})$ in Eq.~\eqref{eq:QReset-indep-2d-in-t-00} and its generating function are given in~\ref{sec:app-indep}.

\begin{figure}[!htbp]
\centering
\includegraphics[scale=0.95]{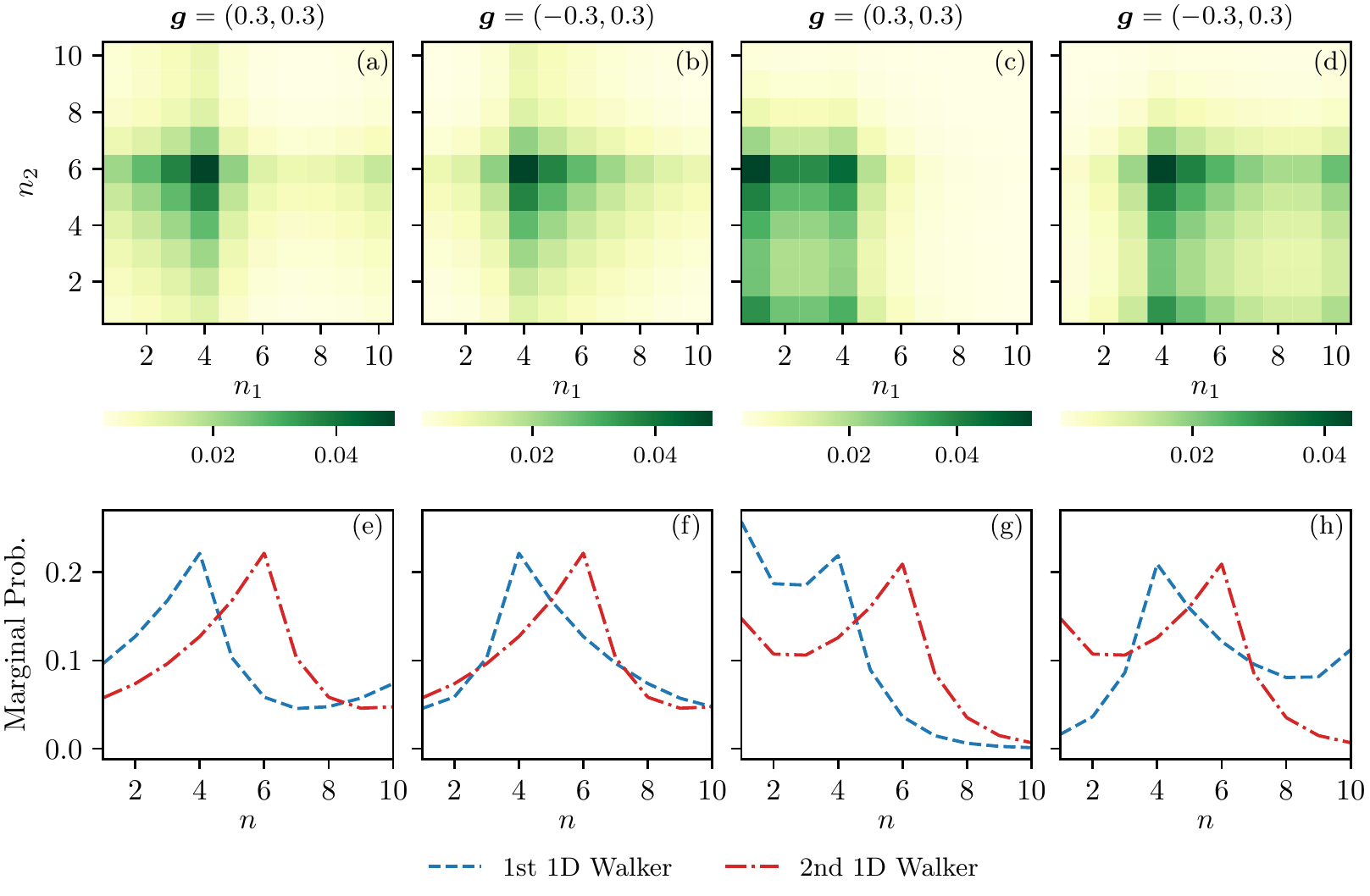} 
\caption{\textbf{2D propagator ${Q}(\bm{n},t|\bm{n_0})$ with independent resetting for different biases at time $t=10^2$.}  The plots are obtained using Eq.~\eqref{eq:QReset-indep-2d-in-t} with $\bm{r}=(0.1,0.1)$. All other parameters are the same as in Fig.~\ref{fig:QReset-simul-joint-2d}. Each panel represents the same quantity as per Fig.~\ref{fig:QReset-simul-joint-2d}.}
\label{fig:QReset-indep-joint-2d}
\end{figure}
The top row of Fig.~\ref{fig:QReset-indep-joint-2d}\, shows the 2D propagator ${{Q}}(\bm{n},t|\bm{n_0})$ with independent resetting in bounded domain.  
Differently from the simultaneous resetting scheme, the occupation probability in the top panels is not only much more distributed over the 2D domain, but also it is prevalent at sites from the sets $\mathfrak{C}$ and $\mathfrak{R}$.
The dependence on the bias of the skewness of the probability is similar to the case of simultaneous resetting.
Note that the marginal probability of a single 1D walker considers all possible occupation states of the other, and thus is independent of the underlying resetting schemes. Consequently, for both resetting schemes, with $r=r_1=r_2$ we get the same marginal probabilities as may be seen from panels (e)--(h) of Figs.~\ref{fig:QReset-simul-joint-2d}\, and \ref{fig:QReset-indep-joint-2d}.

\subsection{Transmission process}
\label{subsec:trans-process}

A transmission event between the 1D walkers can happen only when they are at the same site. In the 2D domain, we refer to these special positions as co-location sites, and denote them by a set of position vectors ${\mathfrak{M}} \equiv \{ \bm{m}_i\}$ with  $\bm{m}_i = (i,i)$, where $i \in [1,2,\ldots, M \equiv \min(N_1,N_2)]$.
The transmission probability at each of these co-location sites $\bm{m}_i$ is given by the constant $\rho$ with $ 0 \leq \rho \leq 1$. 
Consequently, for the 2D walker, each of these co-location sites acts as a partial absorbing site.
These sites represent defects in the lattice and the dynamics, for which the propagator ${\mathcal{P}}(\bm{n},t|\bm{n_0})$ can be obtained from the defect-free propagator $Q(\bm{n},t| \bm{n_0})$ using the defect technique. 
The derivation follows the mathematical steps from Ref.~\cite{giuggioli_spatio-temporal_2022} developed by one of the author of the current paper, and the useful results are summarized in~\ref{app:trans}.
Equation~\eqref{eq:MS-sol-in-z-final} relates the generating function $\widetilde{\mathcal{P}}(\bm{n}, z| \bm{n}_0)$ to the generating function $\widetilde{Q}(\bm{n}, z| \,\bm{n}_0)$ of the defect-free propagator.

The generating function of the survival probability $\widetilde{S}(z|\bm{n_0}) = \sum_{\bm{n}} \widetilde{\mathcal{P}}(\bm{n}, z| \bm{n}_0)$, i.e., the probability of not having been absorbed at any of the sites $\bm{m}_i \in {\mathfrak{M}}$ while starting from site $\bm{n}_0$, 
is given by
\begin{align}
\widetilde{S}(z|\bm{n_0})  = \frac{1}{1-z} -  \frac{\rho}{1-z} \sum_{i=1}^{M}    ~ \frac{\mathcal{D}_i(\rho,z| \bm{n}_0)}{\mathcal{D}(\rho,z)} \, , \label{eq:SP-in-z}
\end{align} 
where the determinant $\mathcal{D}(\rho,z)$ is defined in Eq.~\eqref{eq:D-det}, and the determinant $\mathcal{D}_i(\rho,z | \bm{n}_0 )$ is the same as $\mathcal{D}(\rho,z)$ but with its $i$-th column replaced by the column matrix given in Eq.~\eqref{eq:row_mat}.
When a definite transmission event occurs at time $t$ at any of the co-location sites, the probability $\mathcal{P}(\bm{n},t|\bm{n_0}) $ drops to zero~\cite{kenkre_theory_2014}. The infection probability at time $t$, i.e., the probability of a definite transmission having been occurred at any of the time instant $t' \leq t$ while starting from site $\bm{n}_0$ is given by $\mathcal{I}(\rho,t| \bm{n}_0) \equiv 1- S(\rho,t|\bm{n_0})$~\cite{kenkre_theory_2014}, where  $S(\rho,t|\bm{n_0})$ is the survival probability at time $t$. The generating function of $\mathcal{I}(\rho,t| \bm{n}_0)$ obtained from Eq.~\eqref{eq:SP-in-z} is  
given by
\begin{align}
\widetilde{\mathcal{I}}(\rho,z| \bm{n}_0) = \frac{\rho }{1-z}  \sum_{i=1}^{M}  ~ \frac{\mathcal{D}_i(\rho,z| \bm{n}_0)}{\mathcal{D}(\rho,z)}  \, . \label{eq:IP-in-z}
\end{align}

\begin{figure}[htbp]
\centering
\includegraphics[scale=1.15]{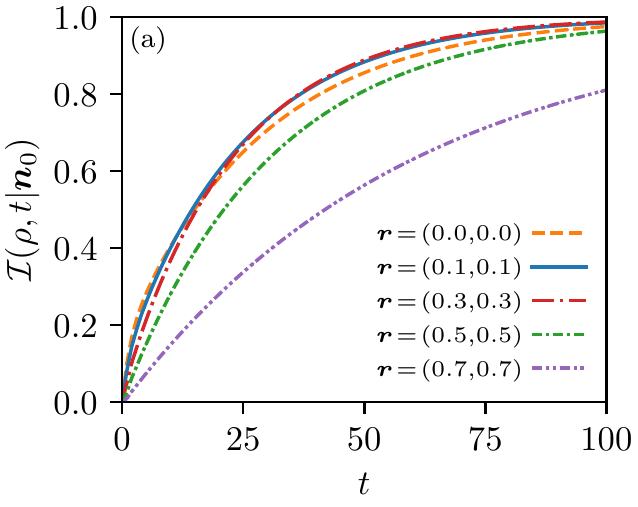} \hspace{5pt}
\includegraphics[scale=1.15]{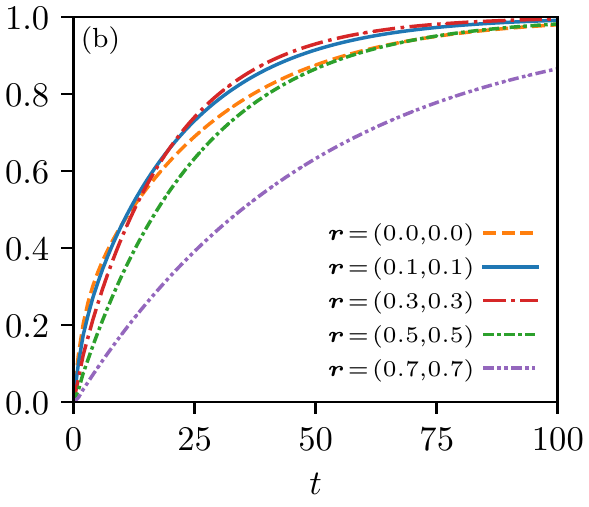} \\[0.5ex]
\includegraphics[scale=1.15]{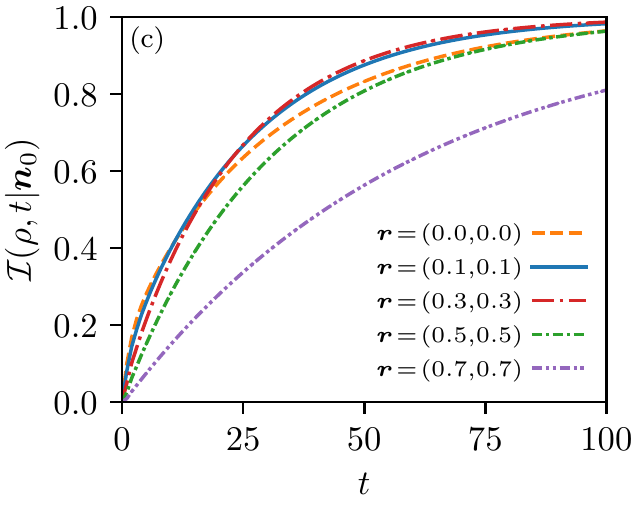} \hspace{5pt}
\includegraphics[scale=1.15]{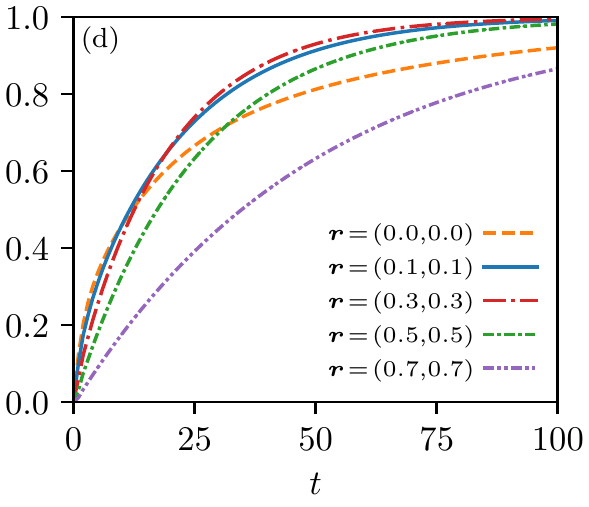} 
\caption{\textbf{Infection probability $\mathcal{I}(\rho,t| \bm{n}_0)$ for independent resetting of the 1D walkers with $\rho=0.5$.} Top row: periodic boundaries. Bottom row: reflecting boundaries. Parameters are the same as in Fig.~\ref{fig:QReset-simul-joint-2d} but with $\bm{g}=(0.1,0.1)$ for panels (a) and (c), and $\bm{g}=(-0.1,0.1)$ for panels (b) and (d). The lines are obtained by numerically inverting Eq.~\eqref{eq:IP-in-z} with the determinants $\mathcal{D}_i(\rho,z| \bm{n}_0), \, \mathcal{D}(\rho,z| \bm{n}_0)$ given by Eqs.~\eqref{eq:D-det},~\eqref{eq:row_mat}, and~\eqref{eq:QReset-indep-2d-in-z}.   }
\label{fig:QReset-IC-indep}
\end{figure}
In Fig.~\ref{fig:QReset-IC-indep}, we have shown the infection probability $\mathcal{I}(\rho,t| \bm{n}_0)$ for different values of $\vect{r}$ in the independent resetting scheme.
For all the plots, keeping the diffusion parameter  $\bm{q}$ and the separation between their resetting sites $|n_{c_1} - n_{c_2}|$ constant, we gradually increase the resetting probabilities $r_1, \, r_2$ starting from $r_1=r_2 = 0$. 
Striking non-monotonic behavior of $\mathcal{I}(\rho,t| \bm{n}_0)$ is apparent from the plots.
For small non-zero values of $r_1$ and $r_2$, it is seen that infection occurs at a faster rate than the reset-free case. However, with further increase in $r_1$ and $ r_2$, infection occurs at a much slower rate because the walkers are more confined in  the neighborhood of their respective resetting sites. 
In panel (d), the reset-free infection rate for large enough values of $t$ is the slowest compared to all other cases. This is owing to the fact that the 1D walkers with opposite biases in reflecting domain always tend to move towards the opposite boundaries and thus the probability of them being at the same site reduces. 
The infection probability shows qualitatively similar non-monotonic behavior also in the simultaneous resetting scheme.
In continuous space-time, for two diffusing walkers confined in quadratic potentials, such non-monotonic infection behavior is observed as the confinement strength is varied~\cite{kenkre_theory_2014}.
Similar non-monotonic behavior appears in the survival probability of Smoluchowski random walkers with trapping in quadratic confinement  potentials~\cite{spendier_reaction-diffusion_2013} as well as in quantum yield in doped molecular crystals~\cite{wolf_energy_1968,powell_singlet_1975,kenkre_memory_2021} and photosynthetic systems~\cite{clayton_photosynthesis_1980} with V-shaped confinement potentials~\cite{chase_analysis_2016}.

\section{Conclusion}
\label{sec:conclu}

In this paper, we have studied the discrete spatio-temporal dynamics of the biased lattice random walk problem with resetting in arbitrary dimensions. We have derived a discrete renewal equation~\eqref{eq:reset-x} and presented a formalism that is applicable to any underlying reset-free Markovian stochastic process. 
We have given working formulae to compute different quantities of the resetting dynamics in terms of the generating function of the underlying reset-free propagator. 
We have then demonstrated the applications of our formalism to the biased lattice walk dynamics with resetting in 1D unbounded domain, for which we have obtained analytical closed-form expressions for the propagator, stationary-state probability, the first-passage probability, the mean first-passage time, and the mean return time.
In presence of a partially absorbing trap, exact expression for the mean absorption time is also presented. We have shown how the discrete space-time expressions reduce to the corresponding relations in the limit of continuous space-time dynamics.
Applications to the biased random walk dynamics in periodic domain and domain limited by reflecting boundaries have also been analysed. In the periodic domain, we have shown the existence of bias-dependent multi-fold non-monotonicity in the first-passage probability as the resetting probability is varied.
Finally, we have applied our formalism to the transmission dynamics of two 1D BLRWs with resetting in bounded domain. This has been done by studying the first-passage statistics of a single 2D BLRW in bounded domain with multiple partially absorbing sites. 
We have found that infection or a definite transmission occurs at a faster rate, when the two 1D BLRWs are subject to resetting with small probabilities, however with increase in these probabilities infection proceeds more slowly. 
A further extension of this work would be the application of our formalism to discrete resetting dynamics of lattice walkers in higher dimensions. Since the propagators for the reset-free lattice-walk dynamics in higher dimensions in finite and infinite domains are now known~\cite{giuggioli_exact_2020,sarvaharman_closed-form_2020}, one can use the working formulae given in Section~\ref{sec:TheResettingModel}\,~to study the corresponding resetting dynamics.

Diffusion in complicated spatial geometries such as comb-like structures are used in various problems, from studying transport in spiny dendrites~\cite{santamaria_anomalous_2006,fedotov_non-markovian_2008} to modelling river basins~\cite{colaiori_analytical_1997,rinaldo_geomorphological_1991}. Studies of more complicated geometries like fractal comb, grid and mesh structures are also important in the context of anomalous diffusion~\cite{iomin_fractional_2018}. 
The continuous space-time model of diffusion with resetting in comb-like geometries are of recent interests~\cite{domazetoski_stochastic_2020,singh_backbone_2021}, and one may apply our formalism to study the discrete resetting dynamics in presence of such spatial geometries. 
Another possible extension could be to discrete heterogeneous diffusive processes with resetting, for which the continuous limits with position-dependent diffusion coefficient have been recently analysed~\cite{sandev_heterogeneous_2022}. One may also study the discrete resetting dynamics in presence of permeable interface, for which a fundamental diffusion equation in the continuous limits has been derived from microscopic description very recently~\cite{kay_diffusion_2022}.

\section{Acknowledgments}
\label{sec:ackno}

We acknowledge funding from the Biotechnology and Biological Sciences Research Council (BBSRC) Grant No. BB/T012196/1. We thank Shamik Gupta for scientific discussions about the derivation of Eq.~\eqref{eq:reset-x}. DD also acknowledges Seeralan Sarvaharman for helpful discussions.
\appendix
\section{Derivation of the renewal equation }
\label{app:derive_renewal_eq}

For the resetting dynamics, we define the probability to find the walker at a site $\bm{n}$ at time $t$ while starting from site $\bm{n}_0$ at time $t=0$ as $Q(\bm{n},t|\bm{n}_0,0)$. The quantity $Q(\bm{n},t|\bm{n}_0,0)$ is expressed in terms of the propagator with no resetting, denoted by $Q_{\mathrm{NR}}(\bm{n},t|\bm{n}_0,0)$, using a sum of terms as follows
\begin{align}
Q(\bm{n},t|\bm{n}_0,0) 
&=  \underbrace{r \, Q_{\mathrm{NR}}({\vect n},t|{\vect n}_c,t)}_{\text{last reset at $t$}}+\underbrace{r \, (1-r) \, Q_{\mathrm{NR}}({\vect n},t|{\vect n}_c,t-1)}_{\text{last reset at $t-1$}}  \nonumber \\ 
&\hskip20pt +\ldots +\underbrace{r \, (1-r)^{t-1} \, Q_{\mathrm{NR}}({\vect n},t|{\vect n}_c,1)}_{\text{last reset at $1$}} + \underbrace{(1-r)^t \, Q_{\mathrm{NR}}({\vect n},t|{\vect n}_0,0)}_{\text{No reset at any of the time steps $1,\ldots,t$}} \nonumber \\
&=r\sum_{t'=0}^{t-1} (1-r)^{t'} Q_{\mathrm{NR}}(\bm{n},t|\bm{n}_c,t-t')+(1-r)^t Q_{\mathrm{NR}}(\bm{n},t|\bm{n}_0,0) \, . 
\label{eq:reset} 
\end{align}
The first term on the right-hand side of Eq.~\eqref{eq:reset} represents the case for the walker to have had the last reset at any of the time steps $1,2,\ldots,t$, while the second represents the case to have had no reset at all during the full evolution up to time $t$.  In the first term, the factor $r(1-r)^{t'}$ is the probability to have had a reset at time step $(t-t')$ and no reset at any of the subsequent time steps $(t-t'+1),\,(t-t'+2),\,\ldots,\, t$, that is for a total of $t'$ time steps.  In the second term, the factor $(1-r)^{t}$ is the probability to have had no reset at steps $1,2,\ldots,t$, that is for a total of $t$ time steps. Note that at time $t=0$, only the second term on the right-hand side of Eq.~\eqref{eq:reset} contributes. Summing Eq.~(\ref{eq:reset}) over all possible values of $\bm{n}$, and using the normalization condition of the reset-free propagator, namely,  $\sum_{\bm{n}} Q_{\mathrm{NR}}(\bm{n},t|\bm{n}_0,0)=1$, we obtain $\sum_{\bm{n}} Q(\bm{n},t|\bm{n}_0,0)=r\sum_{t'=0}^{t-1}(1-r)^{t'}+(1-r)^{t}=r \,[{1-(1-r)^{t}}]/r+(1-r)^{t}=1$. Thus, we check that the resetting propagator $Q(\bm{n},t|\bm{n}_0,0)$ is properly normalized. 

Using the time-translation invariance of the propagator, one may write $Q_{\mathrm{NR}}(\bm{n},t|\bm{n}_c,t-t') = Q_{\mathrm{NR}}(\bm{n},t'|\bm{n}_c,0)$. For brevity in the notation we drop the initial time $t=0$ from the argument of the propagators, so that the quantity $Q(\bm{n},t|\bm{n}_0)$ is to be assumed as the probability to find the walker at site $\bm{n}$ at time $t$ while starting from site $\bm{n}_0$ at time $t=0$. In this lighter notation, Eq.~\eqref{eq:reset} is rewritten as Eq.~\eqref{eq:reset-x}.

\section{Propagators for the BLRW without resetting in $z$-domain}
\label{sec:BLRW}

The generating function of the propagator for the reset-free BLRW in 1D unbounded domain is given by~\cite{sarvaharman_closed-form_2020}
\begin{align}
\widetilde{Q}_{\mathrm{NR}}(n,z|n_0) =  \frac{\eta f^{\frac{n-n_0}{2}} \big[\alpha(z)\big]^{-|n-n_0|}  }{ z q \sinh \left[  \arccosh  \left\{  \frac{\eta -\eta z(1-q)}{z q} \right\}  \right]} \, , \label{eq:defect-free-prop}
\end{align}
where $\eta, \,f$, and $\alpha(z)$ are defined in Eq.~\eqref{eq:eta-f-alpha-def}.
For the reset-free BLRW dynamics in 1D bounded domain $[1,N]$, the time-dependent propagator ${Q}^{}_{\mathrm{NR}}({n},t|n_0)$ is given by~\cite{sarvaharman_closed-form_2020}
\begin{align}
{Q}^{}_{\mathrm{NR}}({n},t|n_0) = \sum_{k = 0}^{N - 1} h^{(N)}_{k} (n, n_0) \Big[ 1+ s^{(N)}_{k} \Big]^t \, ,\label{eq:PsiNR-gen}
\end{align} 
with
\begin{align}
& h^{(N)}_{k} (n, n_0) = \frac{1}{N} ~\ee^{2\ii k \pi (n-n_0)/N}, \label{eq:PsiNR-peri-h}\\
& s^{(N)}_{k} = q\cos\Big( \frac{2k\pi}{N} \Big) + \ii q g \sin\Big( \frac{2k\pi}{N} \Big) -q \, , \label{eq:PsiNR-peri-s}
\end{align}
for periodic boundaries, and 
\begin{align}
& h^{(N)}_{k} (n, n_0) =  \begin{cases} 
      \frac{f^{\frac{n-n_0-1}{2}} \Big[ \sqrt{f} \sin\! \big( \frac{n k \pi }{N} \big) - \, \sin\!\big\{ \frac{(n-1)k\pi }{N} \big\} \Big]  \Big[ \sqrt{f} \sin \!\big( \frac{n_0 k \pi }{N} \big) - \, \sin\! \big\{ \frac{(n_0-1)k\pi }{N} \big\} \Big] }{N \big[\eta \, - \, \cos\! \big( \frac{k\pi}{N} \big)  \big]}\, ; &\hskip5pt  k \neq 0 \, ,\label{eq:PsiNR-refl-h} \\
      \frac{f^{n-1} (1-f)}{1- f ^N} \, ; &\hskip5pt k=0 \, ,
\end{cases} \\
& s^{(N)}_{k} =  \begin{cases} 
      \frac{q}{\eta} \cos\Big( \frac{k\pi}{N} \Big) -q  \, ; &\hskip15pt k \neq 0 \, , \\
      0 \, ;&\hskip15pt k=0 \, ,\label{eq:PsiNR-refl-s}
\end{cases}
\end{align}
for reflecting boundaries, and its generating function is given by
\begin{align}
\widetilde{Q}_{\mathrm{NR}}(n,z|n_0) = \sum_{k = 0}^{N - 1} \frac{h^{(N)}_{k} (n, n_0)}{1-z  \Big[ 1+ s^{(N)}_{k} \Big]} \, . \label{eq:QNR_bounded}
\end{align}

\section{First-passage processes for reset-free BLRW in 1D unbounded domain}
\label{sec:FP-RF-BLRW}

The generating function $\widetilde{\mathcal F}_{\mathrm{NR}}(n,z|n_0) $ of the first-passage probability ${\mathcal  F}_{\mathrm{NR}}(n,t|n_0) $ for the reset-free BLRW in 1D unbounded domain obtained from Eq.~\eqref{eq:defect-free-prop} is given by
\begin{align}
\widetilde{\mathcal  F}_{\mathrm{NR}}(n,z|n_0) = f^{\frac{n-n_0}{2}} \big[\alpha(z) \big]^{-|n-n_0|} \, . \label{eq:FP_no_reset_unbounded}
\end{align}
From Eqs.~\eqref{eq:FP_no_reset_unbounded} and~\eqref{eq:Qreset_MFPT-xx2}, we get $\widetilde{\mathcal  F}_{\mathrm{NR}}(n,1|n_0) = f^{\frac{n-n_0 \pm |n-n_0|}{2}}$, where the positive and the negative part of the  `$\pm$' sign correspond to the cases with $g>0$ and $g< 0$, respectively. 
For an unbiased walker (with $g=0$), we get $\widetilde{\mathcal  F}_{\mathrm{NR}}(n,1|n_0) = 1$.  
Note that for $g>0$, one has $\widetilde{\mathcal  F}_{\mathrm{NR}}(n,1|n_0) \leq 1 $ with the equality holding only if $n \leq n_0$. Similarly, for $g<0$, we get $\widetilde{\mathcal  F}_{\mathrm{NR}}(n,1|n_0) \leq 1 $ with the equality holding only if $n \geq n_0$.  Therefore, in 1D unbounded domain, an unbiased walker  would definitely reach a target at some time, while a biased walker would reach a target only if the target is located in the direction of the bias. In case the target is located in a direction opposite to the bias, there will be trajectories which would never reach the target.
The generating function of the first-return probability $\widetilde{\mathcal  R}_{\mathrm{NR}}({n},z) \equiv  1 - 1 / \widetilde{Q}_{\mathrm{NR}}({n},z|{n})$  is obtained from Eq.~\eqref{eq:defect-free-prop} as
\begin{align}
\widetilde{\mathcal  R}_{\mathrm{NR}}({n},z) = 1 - \frac{z q}{\eta}~ \sinh \bigg[  \arccosh \bigg\{ \frac{\eta - \eta z (1-q)}{z q} \bigg\} \bigg] \, . \label{eq:RP_no_reset_unbounded}
\end{align}
From Eqs.~\eqref{eq:RP_no_reset_unbounded} and~\eqref{eq:Qreset_MFPT-xx2}, we get $ \widetilde{\mathcal  R}_{\mathrm{NR}}({n},1) =  1-q |g|$, which clearly suggests that an unbiased walker (with $g=0$) would definitely return to an initial site at some time. However, for a biased walker (with $g\neq 0$), we have  $\widetilde{\mathcal  R}_{\mathrm{NR}}({n},1) < 1$, and consequently there will be trajectories which would never return to the initial site. 

The MFPT $ T^{\mathrm{NR}}_{n_0 \to n} \equiv [\frac{\partial}{\partial z} \widetilde{ \mathcal  F}_{\mathrm{NR}}(n,z|n_0) ]_{z=1}$ obtained from Eq.~\eqref{eq:FP_no_reset_unbounded} is given by
$ T^{\mathrm{NR}}_{n_0 \to n} = |n-n_0|  \big[  f^{\frac{n-n_0}{2}} (\eta + \sqrt{\eta^2 -1})^{-|n-n_0|} \big]/ q |g| $, 
which clearly diverges for $g=0$. Therefore, although an unbiased walker reaches a target at some time with certainty, the mean time it takes is infinite.  
However, for a biased walker, when it would reach the target with certainty only if the target is located towards the direction of the  bias, the MFPT obtained using Eq.~\eqref{eq:Qreset_MFPT-xx2} is given by
\begin{align}
T^{\mathrm{NR}}_{n_0 \to n} = \frac{ |n-n_0| }{q |g|}  \, . \label{eq:QNR_MFPT}
\end{align}
One may also see from Eq.~\eqref{eq:RP_no_reset_unbounded} that the mean first-return time $R^{\mathrm{NR}}_{ n} \equiv [\frac{\partial}{\partial z} \widetilde{ \mathcal  R}_{\mathrm{NR}}(n,z) ]_{z=1}$ is infinite for an unbiased walker although it is the only case when the first-return probability is normalized.

\section{The 2D propagator for independent resetting}
\label{sec:app-indep}

In presence of independent resetting of the two 1D BLRWs, the time-dependent propagator for the 2D walker is obtained from Eq.~\eqref{eq:QReset-indep-2d-in-t-00}, and is given by
\begin{align}
Q(\bm{n},t|{\bm{n_0}}) &=  \sum_{k_1 = 0}^{N_1 - 1} \sum_{k_2 = 0}^{N_2 - 1}  \Bigg[ \frac{r_1 r_2 \, h^{(N_1)}_{k_1} (n_1, n_{{c_1}}) ~ h^{(N_2)}_{k_2} (n_2, n_{{c_2}})  }{ \big(\gamma^{(N_1)}_{k_1}-1 \big)~  \big(\gamma^{(N_2)}_{k_2}-1\big)} \nonumber \\
&\hskip153pt \times \Big\{ \big({\gamma^{(N_1)}_{k_1}}\big)^{t} \big({\gamma^{(N_2)}_{k_2}}\big)^{t} - \big({\gamma^{(N_1)}_{k_1}}\big)^{t} -\big({\gamma^{(N_2)}_{k_2}}\big)^{t} + 1\Big\} \nonumber \\
&\hskip70pt + \frac{r_2~ h^{(N_1)}_{k_1} (n_1, n_{{0_1}}) ~ h^{(N_2)}_{k_2} (n_2, n_{{c_2}}) }{  \big(\gamma^{(N_2)}_{k_2}-1\big)} ~\Big\{\big({\gamma^{(N_1)}_{k_1}}\big)^{t} \big({\gamma^{(N_2)}_{k_2}}\big)^{t}- \big({\gamma^{(N_1)}_{k_1}}\big)^{t} \Big\}  \nonumber \\
&\hskip70pt + \frac{r_1~ h^{(N_1)}_{k_1} (n_1, n_{{c_1}}) ~ h^{(N_2)}_{k_2} (n_2, n_{{0_2}}) }{   \big(\gamma^{(N_1)}_{k_1}-1\big)} ~\Big\{ \big({\gamma^{(N_1)}_{k_1}}\big)^{t} \big({\gamma^{(N_2)}_{k_2}}\big)^{t} - \big({\gamma^{(N_2)}_{k_2}}\big)^{t} \Big\} \nonumber \\
&\hskip70pt + h^{(N_1)}_{k_1} (n_1, n_{{0_1}}) ~ h^{(N_2)}_{k_2} (n_2, n_{{0_2}})  ~\big({\gamma^{(N_1)}_{k_1}}\big)^{t} \big({\gamma^{(N_2)}_{k_2}}\big)^{t} \Bigg] \, , \label{eq:QReset-indep-2d-in-t} 
\end{align}
where we have $\gamma^{(N_i)}_{k_i} = (1-r_i) [ 1+ s^{(N_i)}_{k_i} ]$. 
For both periodic and reflecting cases, we have $s^{(N_i)}_{0} = 0$, leading to $\lim_{r_i \to 0}~ r_i/(\gamma^{(N_i)}_{k_i}-1) = -1$ for $k_i = 0$  and $ 0$ for $k_i \in[1,N-1] $.
In the limits $r_1 \to 0$ and $ r_2 \to 0$, Eq.~\eqref{eq:QReset-indep-2d-in-t} reduces to the reset-free 2D propagator $Q_{\mathrm{NR}}(\bm{n},t|{\bm{n_0}})$ given in Eq.~\eqref{eq:QNR_2d_simul_gen_in_t}.
The generating function of $Q(\bm{n},t|{\bm{n_0}})$ with independent resetting obtained from Eq.~\eqref{eq:QReset-indep-2d-in-t} yields
\begin{align}
\widetilde{Q}(\bm{n},z|{\bm{n_0}}) &=  \sum_{k_1 = 0}^{N_1 - 1} \sum_{k_2 = 0}^{N_2 - 1} \Bigg[ \frac{r_1 r_2 \, h^{(N_1)}_{k_1} (n_1, n_{{c_1}}) ~ h^{(N_2)}_{k_2} (n_2, n_{{c_2}})  }{ \big(\gamma^{(N_1)}_{k_1}-1 \big)~  \big(\gamma^{(N_2)}_{k_2}-1\big)} \nonumber \\
&\hskip88.43pt \times \Bigg\{ \frac{1}{1-z\,{ \gamma^{(N_1)}_{k_1}} {\gamma^{(N_2)}_{k_2}}} - \frac{1}{1-z\, \gamma^{(N_1)}_{k_1}} -\frac{1}{1-z\,\gamma^{(N_2)}_{k_2}} +\frac{1}{1-z}\Bigg\} \nonumber \\
&\hskip45pt + \frac{r_2~ h^{(N_1)}_{k_1} (n_1, n_{{0_1}}) ~ h^{(N_2)}_{k_2} (n_2, n_{{c_2}}) }{  \big(\gamma^{(N_2)}_{k_2}-1\big)} ~\Bigg\{ \frac{1}{1-z\,{ \gamma^{(N_1)}_{k_1}} {\gamma^{(N_2)}_{k_2}}} - \frac{1}{1-z\,\gamma^{(N_1)}_{k_1}}\Bigg\}  \nonumber \\
&\hskip45pt + \frac{r_1~ h^{(N_1)}_{k_1} (n_1, n_{{c_1}}) ~ h^{(N_2)}_{k_2} (n_2, n_{{0_2}}) }{   \big(\gamma^{(N_1)}_{k_1}-1\big)} ~\Bigg\{ \frac{1}{1-z\,{ \gamma^{(N_1)}_{k_1}} {\gamma^{(N_2)}_{k_2}}} - \frac{1}{1-z\,\gamma^{(N_2)}_{k_2}}\Bigg\} \nonumber \\
&\hskip45pt +    \frac{h^{(N_1)}_{k_1} (n_1, n_{{0_1}}) ~ h^{(N_2)}_{k_2} (n_2, n_{{0_2}})}{1-z \,{ \gamma^{(N_1)}_{k_1}} {\gamma^{(N_2)}_{k_2}} }  \Bigg] \, . \label{eq:QReset-indep-2d-in-z}
\end{align}

\section{The 2D defective propagator in $z$-domain} 
\label{app:trans}

Following Ref.~\cite{giuggioli_spatio-temporal_2022}, here we present the results on the generating function $\widetilde{\mathcal{P}}(\bm{n}, z| \bm{n}_0)$ of the propagator ${\mathcal{P}}(\bm{n},t|\bm{n_0})$ in a 2D domain of size $N_1 \times N_2$ with $M$ partially absorbing sites located at $\bm{m}_i = (i,i)$, where $i \in [1,2,\ldots, M \equiv \min(N_1,N_2)]$. 
For the 2D walker, the absorption probability at these defective sites is given by the constant $\rho~(0\leq \rho\leq 1)$. 
The exact solution for $\widetilde{\mathcal{P}}(\bm{n}, z| \bm{n}_0)$ in terms of the generating function $\widetilde{Q}(\bm{n},z| \bm{n_0})$ of the defect-free propagator $Q(\bm{n},t| \bm{n_0})$ is given by~\cite{giuggioli_spatio-temporal_2022}
\begin{align}
\widetilde{\mathcal{P}}(\bm{n}, z| \bm{n}_0) &=  \widetilde{Q}(\bm{n}, z| \,\bm{n}_0)    -  \rho     \sum_{i=1}^{M} ~ \widetilde{Q}(\bm{n}, z| \,\bm{m}_i) ~ \frac{\mathcal{D}_i(\rho,z| \bm{n}_0)}{\mathcal{D}(\rho,z)}    \, , \label{eq:MS-sol-in-z-final}
\end{align}
where the determinant $\mathcal{D}(\rho,z)$ is given by
\begin{align}
\mathcal{D}(\rho,z) = 
\begin{vmatrix}
1-\rho+\rho \, \widetilde{Q}(\bm{m}_1,z|\bm{m}_1)  & \rho \, \widetilde{Q}(\bm{m}_1,z|\bm{m}_2) & \cdots & \rho \, \widetilde{Q}(\bm{m}_1,z|\bm{m}_M) \\[1ex] 
\rho \, \widetilde{Q}(\bm{m}_2,z|\bm{m}_1) & 1-\rho+\rho \,\widetilde{Q}(\bm{m}_2,z|\bm{m}_2) & \cdots& \rho \, \widetilde{Q}(\bm{m}_2,z|\bm{m}_M) \\ 
\vdots & \vdots & \ddots & \vdots \\ 
\rho \, \widetilde{Q}(\bm{m}_M,z|\bm{m}_1) & \rho \, \widetilde{Q}(\bm{m}_M,z|\bm{m}_2) & \cdots & 1-\rho+\rho \, \widetilde{Q}(\bm{m}_M,z|\bm{m}_M)  
\end{vmatrix} \, , \label{eq:D-det}
\end{align}
while the determinant $\mathcal{D}_i(\rho,z | \bm{n}_0 )$ is the same as $\mathcal{D}(\rho,z)$ but with its $i$-th column replaced by the column matrix
\begin{align}
C_i = \begin{pmatrix}
\widetilde{Q}(\bm{m}_1,z|\bm{n}_0) && \widetilde{Q}(\bm{m}_2,z|\bm{n}_0) && \cdots && \widetilde{Q}(\bm{m}_M,z|\bm{n}_0)
\end{pmatrix}^{\mathsf{T}} \, . \label{eq:row_mat}
\end{align}

\section*{References}

\bibliographystyle{unsrt} 
\bibliography{library}

\end{document}